\documentclass[journal]{IEEEtran}

\usepackage[margin=0.5in]{geometry}
\usepackage{textcomp}
\usepackage{pgfplots}
\usepackage{amsmath}
\usepackage{multicol}
\usepackage{graphicx}
\graphicspath{{Figures/}}
\usepackage{epstopdf}
\usepackage{acronym}
\usepackage{amsmath}
\usepackage{amsfonts}
\usepackage{color, colortbl}
\usepackage{subfigure}
\usepackage[english]{babel}
\usepackage{blindtext}
\usepackage{algorithm} %format of the algorithm
\usepackage{algorithmic} %format of the algorithm
\usepackage{multirow}
\usepackage{color}
\usepackage{subfigure}
\usepackage{balance}
\usepackage{caption}

\begin{document}

\newacro{3GPP}{third generation partnership project}
\newacro{4G}{4{th} generation}
\newacro{5G}{5{th} generation}

\newacro{Adam}{adaptive moment estimation}
\newacro{ADC}{analogue-to-digital converter}
\newacro{AED}{accumulated euclidean distance}
\newacro{AGC}{automatic gain control}
\newacro{AI}{artificial intelligence}
\newacro{AMB}{adaptive multi-band}
\newacro{AMB-SEFDM}{adaptive multi-band SEFDM}
\newacro{AN}{artificial noise}
\newacro{ANN}{artificial neural network}
\newacro{AoA}{angle of arrival}
\newacro{ASE}{amplified spontaneous emission}
\newacro{ASIC}{application specific integrated circuit}
\newacro{AWG}{arbitrary waveform generator}
\newacro{AWGN}{additive white Gaussian noise}
\newacro{A/D}{analog-to-digital}

\newacro{B2B}{back-to-back}
\newacro{BCF}{bandwidth compression factor}
\newacro{BCJR}{Bahl-Cocke-Jelinek-Raviv}
\newacro{BDM}{bit division multiplexing}
\newacro{BED}{block efficient detector}
\newacro{BER}{bit error rate}
\newacro{Block-SEFDM}{block-spectrally efficient frequency division multiplexing}
\newacro{BLER}{block error rate}
\newacro{BPSK}{binary phase shift keying}
\newacro{BS}{base station}
\newacro{BSS}{best solution selector}
\newacro{BT}{British Telecom}
\newacro{BU}{butterfly unit}

\newacro{CapEx}{capital expenditure}
\newacro{CA}{carrier aggregation}
\newacro{CBS}{central base station}
\newacro{CC}{component carriers}
\newacro{CCDF}{complementary cumulative distribution function}
\newacro{CCE}{control channel element}
\newacro{CCs}{component carriers}
\newacro{CD}{chromatic dispersion}
\newacro{CDF}{cumulative distribution function}
\newacro{CDI}{channel distortion information}
\newacro{CDMA}{code division multiple access}
\newacro{CI}{constructive interference}
\newacro{CIR}{carrier-to-interference power ratio}
\newacro{CMOS}{complementary metal-oxide-semiconductor}
\newacro{CNN}{convolutional neural network}
\newacro{CoMP}{coordinated multiple point}
\newacro{CO-SEFDM}{coherent optical-SEFDM}
\newacro{CP}{cyclic prefix}
\newacro{CPE}{common phase error}
\newacro{CRVD}{conventional real valued decomposition}
\newacro{CR}{cognitive radio}
\newacro{CRC}{cyclic redundancy check}
\newacro{CS}{central station}
\newacro{CSI}{channel state information}
\newacro{CSIT}{channel state information at the transmitter}
\newacro{CSPR}{carrier to signal power ratio}
\newacro{CW}{continuous-wave}
\newacro{CWT}{continuous wavelet transform}
\newacro{C-RAN}{cloud-radio access networks}

\newacro{DAC}{digital-to-analogue converter}
\newacro{DBP}{digital backward propagation}
\newacro{DC}{direct current}
\newacro{DCGAN}{deep convolutional generative adversarial network}
\newacro{DCI}{downlink control information}
\newacro{DCT}{discrete cosine transform}
\newacro{DDC}{digital down-conversion}
\newacro{DDO-OFDM}{directed detection optical-OFDM}
\newacro{DDO-OFDM}{direct detection optical-OFDM}
\newacro{DDO-SEFDM}{directed detection optical-SEFDM}
\newacro{DFB}{distributed feedback}
\newacro{DFDMA}{distributed FDMA}
\newacro{DFT}{discrete Fourier transform}
\newacro{DFrFT}{discrete fractional Fourier transform}
\newacro{DL}{deep learning}
\newacro{DMA}{direct memory access}
\newacro{DMRS}{demodulation reference signal}
\newacro{DoF}{degree of freedom}
\newacro{DOFDM}{dense orthogonal frequency division multiplexing}
\newacro{DP}{dual polarization}
\newacro{DPC}{dirty paper coding}
\newacro{DSB}{double sideband}
\newacro{DSL}{digital subscriber line}
\newacro{DSP}{digital signal processors}
\newacro{DSSS}{direct sequence spread spectrum}
\newacro{DT}{decision tree}
\newacro{DVB}{digital video broadcast}
\newacro{DWDM}{dense wavelength division multiplexing}
\newacro{DWT}{discrete wavelet transform}
\newacro{D/A}{digital-to-analog}

\newacro{ECC}{error correcting codes}
\newacro{ECL}{external-cavity laser}
\newacro{ECOC}{error-correcting output codes}
\newacro{EDFA}{erbium doped fiber amplifier}
\newacro{EE}{energy efficiency}
\newacro{eMBB}{enhanced mobile broadband}
\newacro{eNB-IoT}{enhanced NB-IoT}
\newacro{EPA}{extended pedestrian A}
\newacro{EVM}{error vector magnitude}

\newacro{Fast-OFDM}{fast-orthogonal frequency division multiplexing}
\newacro{FBMC}{filter bank multicarrier }
\newacro{FCE}{full channel estimation}
\newacro{FD}{fixed detector}
\newacro{FDD}{frequency division duplexing}
\newacro{FDM}{frequency division multiplexing}
\newacro{FDMA}{frequency division multiple access}
\newacro{FE}{full expansion}
\newacro{FEC}{forward error correction}
\newacro{FEXT}{far-end crosstalk}
\newacro{FF}{flip-flop}
\newacro{FFT}{fast Fourier transform}
\newacro{FFTW}{Fastest Fourier Transform in the West}
\newacro{FHSS}{frequency-hopping spread spectrum}
\newacro{FIFO}{first in first out}
\newacro{FMCW}{frequency-modulated continuous wave}
\newacro{F-OFDM}{filtered-orthogonal frequency division multiplexing}
\newacro{FPGA}{field programmable gate array}
\newacro{FrFT}{fractional Fourier transform}
\newacro{FSD}{fixed sphere decoding}
\newacro{FSD-MNSF}{FSD-modified-non-sort-free}
\newacro{FSD-NSF}{FSD-non-sort-free}
\newacro{FSD-SF}{FSD-sort-free}
\newacro{FSK}{frequency shift keying}
\newacro{FTN}{faster than Nyquist}
\newacro{FTTB}{fiber to the building}
\newacro{FTTC}{fiber to the cabinet}
\newacro{FTTdp}{fiber to the distribution point}
\newacro{FTTH}{fiber to the home}

\newacro{GAN}{generative adversarial network}
\newacro{GB}{guard band}
\newacro{GFDM}{generalized frequency division multiplexing}
\newacro{GNN}{graph neural networks}
\newacro{GPU}{graphics processing unit}
\newacro{GSM}{global system for mobile communication}
\newacro{GUI}{graphical user interface}

\newacro{HARQ}{hybrid automatic repeat request}
\newacro{HC-MCM}{high compaction multi-carrier communication}
\newacro{HPA}{high power amplifier}

\newacro{IC}{integrated circuit}
\newacro{ICI}{inter carrier interference}
\newacro{ICT}{Information and communications technology}
\newacro{ID}{iterative detection}
\newacro{IDCT}{inverse discrete cosine transform}
\newacro{IDFT}{inverse discrete Fourier transform}
\newacro{IDFrFT}{inverse discrete fractional Fourier transform}
\newacro{ID-FSD}{iterative detection-FSD}
\newacro{ID-SD}{ID-sphere decoding}
\newacro{IF}{intermediate frequency}
\newacro{IFFT}{inverse fast Fourier transform}
\newacro{IFrFT}{inverse fractional Fourier transform}
\newacro{IIoT}{industrial Internet of things}
\newacro{IM}{index modulation}
\newacro{IMD}{intermodulation distortion}
\newacro{INOFS}{inverse non-orthogonal frequency shaping}
\newacro{IoT}{Internet of things}
\newacro{IOTA}{isotropic orthogonal transform algorithm}
\newacro{IP}{intellectual property}
\newacro{IR}{infrared}
\newacro{ISAC}{integrated sensing and communication}
\newacro{ISAR}{inverse synthetic aperture radar}
\newacro{ISC}{interference self cancellation}
\newacro{ISI}{inter symbol interference}
\newacro{ISM}{industrial, scientific and medical}
\newacro{ISTA}{iterative shrinkage and thresholding}
\newacro{IWAI}{integrated waveform and intelligence}

\newacro{KNN}{k-nearest neighbours}

\newacro{LDPC}{low density parity check}
\newacro{LFDMA}{localized FDMA}
\newacro{LLR}{log-likelihood ratio}
\newacro{LNA}{low noise amplifier}
\newacro{LO}{local oscillator}
\newacro{LOS}{line-of-sight}
\newacro{LPWAN}{low power wide area network}
\newacro{LS}{least square}
\newacro{LSTM}{long short-term memory}
\newacro{LTE}{long term evolution}
\newacro{LTE-Advanced}{long term evolution-advanced}
\newacro{LUT}{look-up table}

\newacro{MA}{multiple access}
\newacro{MAC}{media access control}
\newacro{MAMB}{mixed adaptive multi-band}
\newacro{MAMB-SEFDM}{mixed adaptive multi-band SEFDM}
\newacro{MASK}{m-ary amplitude shift keying}
\newacro{MB}{multi-band}
\newacro{MB-SEFDM}{multi-band SEFDM}
\newacro{MCM}{multi-carrier modulation}
\newacro{MC-CDMA}{multi-carrier code division multiple access}
\newacro{MCS}{modulation and coding scheme}
\newacro{MF}{matched filter}
\newacro{MIMO}{multiple input multiple output}
\newacro{ML}{maximum likelihood}
\newacro{MLSD}{maximum likelihood sequence detection}
\newacro{MMF}{multi-mode fiber}
\newacro{MMSE}{minimum mean squared error}
\newacro{mMTC}{massive machine-type communication}
\newacro{MNSF}{modified-non-sort-free}
\newacro{MOFDM}{masked-OFDM}
\newacro{MRVD}{modified real valued decomposition}
\newacro{MS}{mobile station}
\newacro{MSE}{mean squared error}
\newacro{MTC}{machine-type communication}
\newacro{MUI}{multi-user interference}
\newacro{MUSA}{multi-user shared access}
\newacro{MU-MIMO}{multi-user multiple-input multiple-output}
\newacro{MZM}{Mach-Zehnder modulator}
\newacro{M2M}{machine to machine}

\newacro{NB-IoT}{narrowband IoT}
\newacro{NB}{naive Bayesian}
\newacro{NDFF}{National Dark Fiber Facility}
\newacro{NEXT}{near-end crosstalk}
\newacro{NFV}{network function virtualization}
\newacro{NG-IoT}{next generation IoT}
\newacro{NLOS}{non-line-of-sight}
\newacro{NLSE}{nonlinear Schrödinger equation}
\newacro{NN}{neural network}
\newacro{NOFDM}{non-orthogonal frequency division multiplexing}
\newacro{NOMA}{non-orthogonal multiple access}
\newacro{NoFDMA}{non-orthogonal frequency division multiple access}
\newacro{NOFS}{non-orthogonal frequency spacing}
\newacro{NP}{non-polynomial}
\newacro{NR}{new radio}
\newacro{NSF}{non-sort-free}
\newacro{NWDM}{Nyquist wavelength division multiplexing }
\newacro{Nyquist-SEFDM}{Nyquist-spectrally efficient frequency division multiplexing}

\newacro{OBM-OFDM}{orthogonal band multiplexed OFDM}
\newacro{ODDM}{orthogonal delay-Doppler division multiplexing}
\newacro{OF}{optical filter}
\newacro{OFDM}{orthogonal frequency division multiplexing}
\newacro{OFDMA}{orthogonal frequency division multiple access}
\newacro{OMA}{orthogonal multiple access}
\newacro{OpEx}{operating expenditure}
\newacro{OPM}{optical performance monitoring}
\newacro{OQAM}{offset-QAM}
\newacro{OSI}{open systems interconnection}
\newacro{OSNR}{optical signal-to-noise ratio}
\newacro{OSSB}{optical single sideband}
\newacro{OTA}{over-the-air}
\newacro{OTFS}{orthogonal time frequency space}
\newacro{Ov-FDM}{Overlapped FDM}
\newacro{O-SEFDM}{optical-spectrally efficient frequency division multiplexing}
\newacro{O-FOFDM}{optical-fast orthogonal frequency division multiplexing}
\newacro{O-OFDM}{optical-orthogonal frequency division multiplexing}
\newacro{O-CDMA}{optical-code division multiple access}

\newacro{PA}{power amplifier}
\newacro{PAPR}{peak-to-average power ratio}
\newacro{PCA}{principal component analysis}
\newacro{PCE}{partial channel estimation}
\newacro{PD}{photodiode}
\newacro{PDCCH}{physical downlink control channel}
\newacro{PDF}{probability density function}
\newacro{PDP}{power delay profile}
\newacro{PDMA}{polarisation division multiple access}
\newacro{PDM-OFDM}{polarization-division multiplexing-OFDM}
\newacro{PDM-SEFDM}{polarization-division multiplexing-SEFDM}
\newacro{PDSCH}{physical downlink shared channel}
\newacro{PE}{processing element}
\newacro{PED}{partial Euclidean distance}
\newacro{PLA}{physical layer authentication}
\newacro{PLS}{physical layer security}
\newacro{PMD}{polarization mode dispersion}
\newacro{PON}{passive optical network}
\newacro{PPM}{parts per million}
\newacro{PRB}{physical resource block}
\newacro{PSD}{power spectral density}
\newacro{PSK}{pre-shared key}
\newacro{PSNR}{peak signal-to-noise ratio}
\newacro{PSS}{primary synchronization signal}
\newacro{PU}{primary user}
\newacro{PXI}{PCI extensions for instrumentation}
\newacro{P/S}{parallel-to-serial}

\newacro{QAM}{quadrature amplitude modulation}
\newacro{QKD}{quantum key distribution}
\newacro{QoS}{quality of service}
\newacro{QPSK}{quadrature phase-shift keying}
\newacro{QRNG}{quantum random number generation}

\newacro{RAUs}{remote antenna units}
\newacro{RAT}{radio access technology}
\newacro{RBF}{radial basis function}
\newacro{RBW}{resolution bandwidth}
\newacro{ReLU}{rectified linear units}
\newacro{RF}{radio frequency}
\newacro{RMS}{root mean square}
\newacro{RMSE}{root mean square error}
\newacro{RMSProp}{root mean square propagation}
\newacro{RNTI}{radio network temporary identifier}
\newacro{RoF}{radio-over-fiber}
\newacro{ROM}{read only memory}
\newacro{RRC}{root raised cosine}
\newacro{RC}{raised cosine}
\newacro{RSC}{recursive systematic convolutional}
\newacro{RSSI}{received signal strength indicator}
\newacro{RTL}{register transfer level}
\newacro{RVD}{real valued decomposition}

\newacro{SB-SEFDM}{single-band SEFDM}
\newacro{ScIR}{sub-carrier to interference ratio}
\newacro{SCMA}{sparse code multiple access}
\newacro{SC-NOFS}{single-carrier non-orthogonal frequency shaping}
\newacro{SC-OFDM}{single-carrier orthogonal frequency division multiplexing}
\newacro{SC-FDMA}{single-carrier frequency division multiple access}
\newacro{SC-SEFDMA}{single-carrier spectrally efficient frequency division multiple access}
\newacro{SD}{sphere decoding}
\newacro{SDM}{space division multiplexing}
\newacro{SDMA}{space division multiple access}
\newacro{SDN}{software-defined network}
\newacro{SDP}{semidefinite programming}
\newacro{SDR}{software-defined radio}
\newacro{SE}{spectral efficiency}
\newacro{SEFDM}{spectrally efficient frequency division multiplexing}
\newacro{SEFDMA}{spectrally efficient frequency division multiple access} 
\newacro{SF}{sort-free}
\newacro{SFCW}{stepped-frequency continuous wave}
\newacro{SGD}{stochastic gradient descent}
\newacro{SGDM}{stochastic gradient descent with momentum}
\newacro{SIC}{successive interference cancellation}
\newacro{SiGe}{silicon-germanium}
\newacro{SINR}{signal-to-interference-plus-noise ratio}
\newacro{SIR}{signal-to-interference ratio}
\newacro{SISO}{single-input single-output}
\newacro{SLM}{spatial light modulator}
\newacro{SMF}{single mode fiber}
\newacro{SNR}{signal-to-noise ratio}
\newacro{SP}{shortest-path}
\newacro{SPSC}{symbol per signal class}
\newacro{SPM}{self-phase modulation}
\newacro{SRS}{sounding reference signal}
\newacro{SSB}{single-sideband}
\newacro{SSBI}{signal-signal beat interference}
\newacro{SSFM}{split-step Fourier method}
\newacro{SSMF}{standard single mode fiber}
\newacro{STBC}{space time block coding}
\newacro{STFT}{short time Fourier transform}
\newacro{STC}{space time coding}
\newacro{STO}{symbol timing offset}
\newacro{SU}{secondary user}
\newacro{SVD}{singular value decomposition}
\newacro{SVM}{support vector machine}
\newacro{SVR}{singular value reconstruction}
\newacro{S/P}{serial-to-parallel}

\newacro{TDD}{time division duplexing}
\newacro{TDMA}{time division multiple access }
\newacro{TDM}{time division multiplexing}
\newacro{TFP}{time frequency packing}
\newacro{THP}{Tomlinson-Harashima precoding}
\newacro{TOFDM}{truncated OFDM}
\newacro{TSPSC}{training symbols per signal class}
\newacro{TSVD}{truncated singular value decomposition}
\newacro{TSVD-FSD}{truncated singular value decomposition-fixed sphere decoding}
\newacro{TTI}{transmission time interval}

\newacro{UAV}{unmanned aerial vehicle}
\newacro{UCR}{user compression ratio}
\newacro{UE}{user equipment}
\newacro{UFMC}{universal-filtered multi-carrier}
\newacro{ULA}{uniform linear array}
\newacro{UMTS}{universal mobile telecommunications service}
\newacro{URLLC}{ultra-reliable low-latency communication}
\newacro{USRP}{universal software radio peripheral}
\newacro{UWB}{ultra-wideband}

\newacro{VDSL}{very-high-bit-rate digital subscriber line}
\newacro{VDSL2}{very-high-bit-rate digital subscriber line 2}
\newacro{VHDL}{very high speed integrated circuit hardware description language}
\newacro{VLC}{visible light communication}
\newacro{VLSI}{very large scale integration}
\newacro{VOA}{variable optical attenuator}
\newacro{VP}{vector perturbation}
\newacro{VSSB-OFDM}{virtual single-sideband OFDM}
\newacro{V2V}{vehicle-to-vehicle}

\newacro{WAN}{wide area network}
\newacro{WCDMA}{wideband code division multiple access}
\newacro{WDM}{wavelength division multiplexing}
\newacro{WDP}{waveform-defined privacy}
\newacro{WDS}{waveform-defined security}
\newacro{WiFi}{wireless fidelity}
\newacro{WiGig}{Wireless Gigabit Alliance}
\newacro{WiMAX}{Worldwide interoperability for Microwave Access}
\newacro{WLAN}{wireless local area network}
\newacro{WSS}{wavelength selective switch}

\newacro{XPM}{cross-phase modulation}

\newacro{ZF}{zero forcing}
\newacro{ZP}{zero padding}

%\newacro{CIR}{channel impulse response}

\title{OFDM-Standard Compatible SC-NOFS Waveforms for Low-Latency and Jitter-Tolerance\\Industrial IoT Communications}

\author{{Tongyang Xu,~\IEEEmembership{Member,~IEEE}}, {Shuangyang Li,~\IEEEmembership{Member,~IEEE}}, {Jinhong Yuan,~\IEEEmembership{Fellow,~IEEE}}
\thanks{
This work was supported by the UK Engineering and Physical Sciences Research Council (EPSRC) under Grant EP/Y000315/1.

T. Xu is with the School of Engineering, Newcastle University, Newcastle upon Tyne, NE1 7RU, U.K. (e-mail: tongyang.xu@newcastle.ac.uk). 

S. Li is with the Department of Electrical Engineering and Computer Science, Technical University of Berlin, Berlin 10587, Germany (e-mail: shuangyang.li@tu-berlin.de)

J. Yuan is with the School of Electrical Engineering and Telecommunications, University of New South Wales, Sydney, NSW 2052, Australia (e-mail: j.yuan@unsw.edu.au).
}}

\maketitle

\begin{abstract}

Traditional communications focus on regular and orthogonal signal waveforms for simplified signal processing and improved spectral efficiency. In contrast, the next-generation communications would aim for irregular and non-orthogonal signal waveforms to introduce new capabilities. This work proposes a spectrally efficient irregular Sinc (irSinc) shaping technique, revisiting the traditional Sinc back to 1924, with the aim of enhancing performance in industrial Internet of things (IIoT). In time-critical IIoT applications, low-latency and time-jitter tolerance are two critical factors that significantly impact the performance and reliability. Recognizing the inevitability of latency and jitter in practice, this work aims to propose a waveform technique to mitigate these effects via reducing latency and enhancing the system robustness under time jitter effects. The utilization of irSinc yields a signal with increased spectral efficiency without sacrificing error performance. Integrating the irSinc in a two-stage framework, a single-carrier non-orthogonal frequency shaping (SC-NOFS)  waveform is developed, showcasing perfect compatibility with 5G standards, enabling the direct integration of irSinc in existing industrial IoT setups. Through 5G standard signal configuration, our signal achieves faster data transmission within the same spectral bandwidth. Hardware experiments validate an 18\% saving in timing resources, leading to either reduced latency or enhanced jitter tolerance.

\end{abstract}


\section{Introduction} \label{sec:introduction}

\IEEEPARstart{T}{he} advancement of \ac{IIoT} \cite{IIoT_survey_2014,IIoT_survey_2017,IIoT_survey_2022} has enabled various time-critical applications in factory automation, power system control, and transportation. However, the practical implementation of IIoT presents challenges \cite{IIoT_mag_2022,IIoT_survey_2018}, particularly concerning latency and jitter. Latency, or the delay in data transmission, is a critical concern in applications where real-time responsiveness is critical, such as robotic control or automated manufacturing. Jitter, the variability in data packet delivery times, can disrupt the synchronization of processes, impacting the reliability of industrial systems. The need for jitter-tolerance, low-latency networks, adherence to standards, and careful consideration of the specific requirements of industrial applications are essential for successful IIoT implementations.

The predominant approaches currently focus on optimizing \ac{MAC} layer protocols through methods like time-sensitive networking (TSN) \cite{IIoT_TSN_mag_2018}, deterministic networking (DetNet) \cite{IIoT_DetNet_2019}, and time synchronous channel hopping (TSCH) \cite{IIoT_TSCH_SJ_2013}. However, the options for enhancing physical layer (PHY) techniques are restricted, given that incorporating a new PHY technique requires adjustments to the existing PHY layer standards, posing a challenge that is impractical to overcome. Consequently, existing PHY layer techniques are mainly focused on adjusting communication system configurations, such as incorporating mmWave frequencies and \ac{MIMO} technology. Nevertheless, the utilization of mmWave frequencies introduces significant challenges, including signal attenuation, high power consumption, and increased costs. The use of \ac{MIMO} architecture adds complexity to antenna design, increased power consumption, and extra latency stemming from \ac{CSI} feedback processing. Hence, there is a need to innovate in PHY layer techniques to address latency and jitter challenges.

\begin{figure*}[t!]
\begin{center}
\includegraphics[scale=0.41]{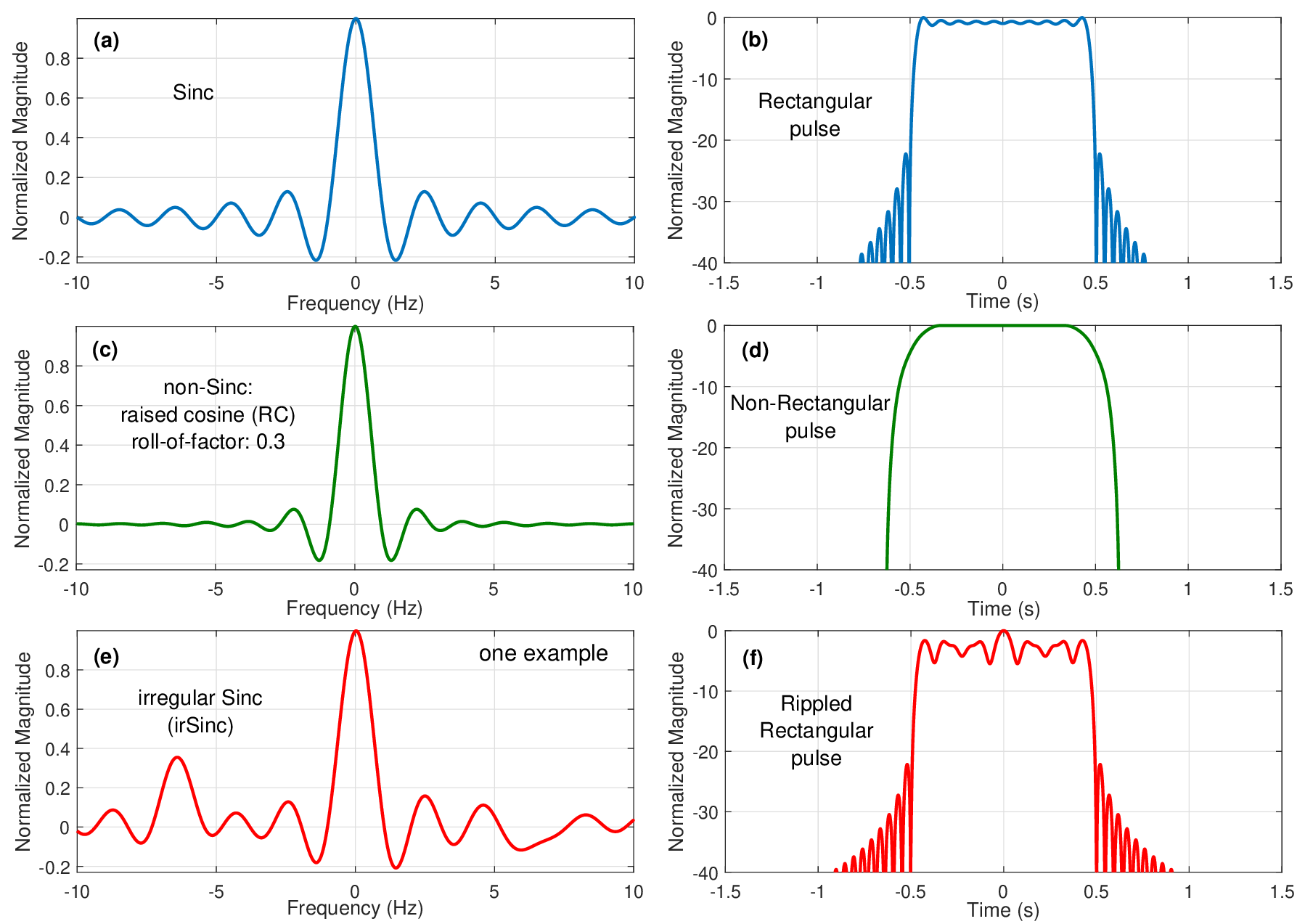}
\end{center}
\caption{Time and frequency characteristics for Sinc, raised cosine (RC), and irSinc patterns. }
\label{Fig:shape_subcarrier_NOFS_Sinc_RRC_irSinc_pure}
\end{figure*}

Waveform design is an important research topic in communications dated back to 1924 \cite{Nyquist_1924} where Nyquist discovered a signal shape to be used in band limited telegraph systems. Furthermore, in 1928 \cite{Nyquist_1928}, Nyquist defined the well-known Nyquist rate that provides an interference-free signal design principle being used in all communication systems. Building upon the communication principles established by Nyquist, in 1949 \cite{Shannon_weaver_1949}, Shannon's capacity theory was proposed, shaping the communication research from that point onward. The \ac{OFDM} was developed in 1958 \cite{OFDM1958} and became the foundation for modern communication systems. In 4G \ac{LTE}, \ac{OFDM} was standardized due to its enhanced spectral efficiency and simple signal processing. To further enhance spectral efficiency and lead 5G, many waveform candidates have been proposed with non-orthogonal signal features. Representative signals are \ac{SEFDM} \cite{TongyangTVT2017}, \ac{FTN} \cite{FTN_IEEE_access_2021}, \ac{FBMC} \cite{FBMC2011_magazine}, and \ac{GFDM} \cite{GFDM_trans}. They all improve spectral efficiency while intentionally introduce \ac{ICI} and \ac{ISI}, leading to complicated signal generation, channel estimation, and signal detection~\cite{TongyangCL2013,Anderson2013FTN,Shuangyang_FTN1,Shuangyang_FTN2}. The sophisticated signal processing prevents the inclusion of the aforementioned non-orthogonal signals in 5G, thus maintaining OFDM as the standardized waveform format.

For next-generation 6G \cite{6G_WC_2020,6G_Network_2020,6G_CM_2020, book_6G}, diverse services are expected to appear and new functional signal waveforms are needed. The next generation waveform (NGW) can be roughly characterized by orthogonal and non-orthogonal signal formats. The orthogonal waveform representatives may be the recently proposed multicarrier waveform based on delay-Doppler processing, e.g.,~\ac{OTFS} \cite{OTFS_WCNC2017} and \ac{ODDM} \cite{ODDM_TWC_2022}. These waveforms are designed carefully to deal with delay-Doppler effect in doubly-selective wireless communications. The non-orthogonal representative, on the other hand, may be the \ac{NOFS} \cite{Tongyang_Nature_2023}, whose main objective is to enhance spectral efficiency beyond the Mazo limit \cite{Mazo1975} via compressing spectral or timing utilization. Both orthogonal and non-orthogonal signal waveforms bring apparent benefits, but they still have challenges in increased signal processing complexity due to the break of traditional signal transmission framework. It is expected that future 6G might still use existing hardware infrastructure, driven by substantial prior investments \cite{NGMN_6G_statement, 6G_industry_need_2023} and a commitment to sustainability. In this case, opting for a fundamentally new signal waveform may not be the primary choice, as it would require the investment of new hardware into the existing system. Therefore, a more suitable approach would maintain the current over-the-air signal waveform format, \ac{OFDM}, and implement minor modifications from other dimensions.

To simplify signal processing and have a better integration of non-orthogonal signal waveforms in existing systems, this work aims to develop a new signal framework, termed \ac{SC-NOFS}, which reserves its increased spectral efficiency advantage and provides compatibility with existing 3GPP standard. Similar to traditional \ac{SC-OFDM}, our proposed SC-NOFS requires two-stage processing. The first stage is our proposed non-orthogonal signal processing and the second stage employs traditional \ac{IFFT} signal generation. This system architecture not only enhances spectral efficiency but also reserves the benefit of OFDM signal characteristics over the air leading to simplified signal detection and channel estimation. In addition, this system architecture provides flexibility in the first stage since any user defined signal patterns can be integrated here.

\begin{figure*}[t!]
\begin{center}
\includegraphics[scale=0.48]{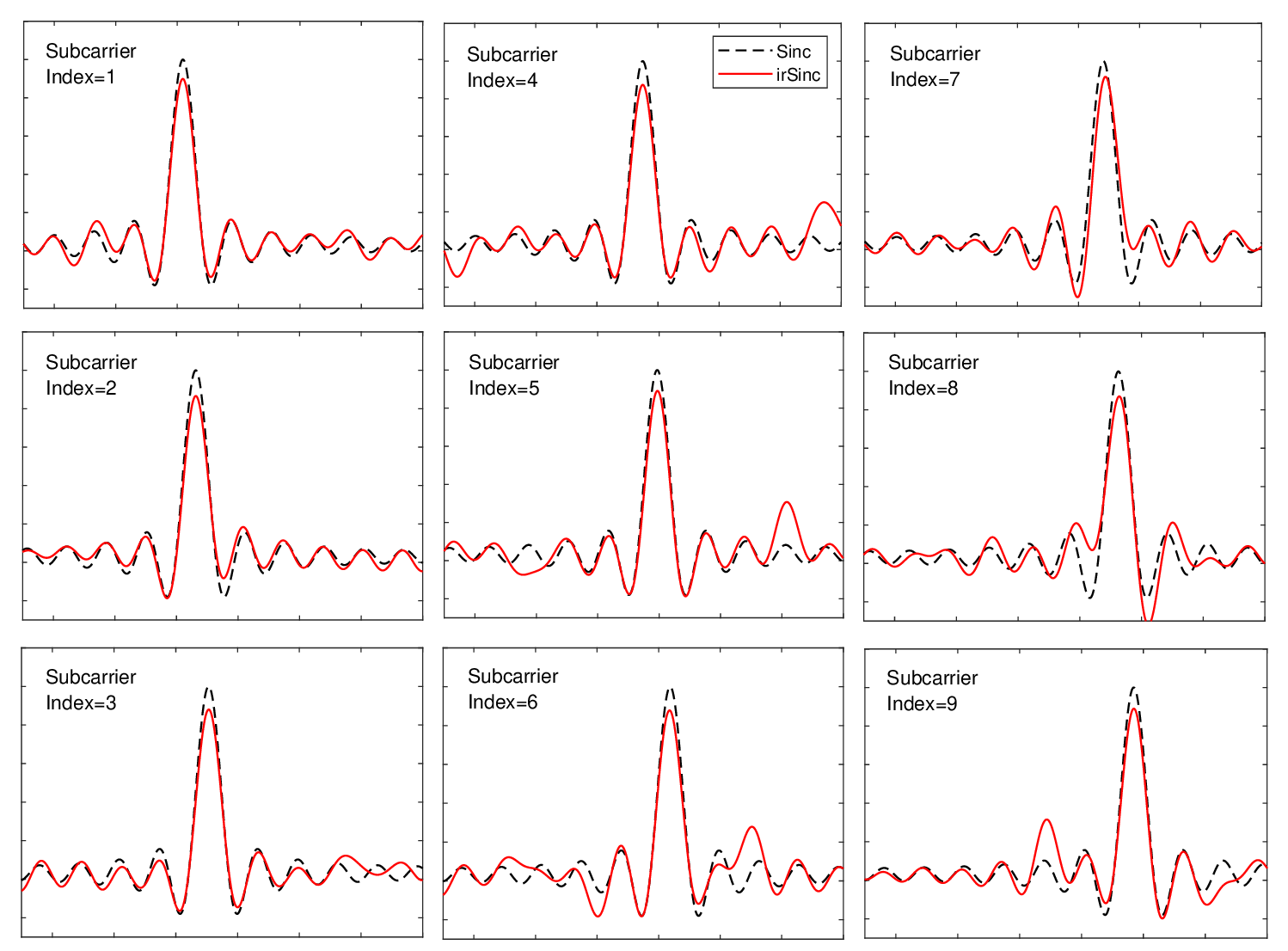}
\end{center}
\caption{irSinc pattern at each sub-carrier.}
\label{Fig:NOFS_irSinc_subcarrier}
\end{figure*}

The contributions of this work are as follow:

\begin{itemize}

\item{ {An OFDM-standard compatible \ac{SC-NOFS} signal is proposed. The two-stage bandwidth compressed waveform framework offers flexibility in designing user-defined signal patterns in the first stage, while maintaining compatibility with the OFDM waveform standard in the second stage. The first stage enhances spectral efficiency, while the second stage ensures signal orthogonality, simplifying channel estimation and equalization. }}

\item{{Reduced signal processing complexity in \ac{SC-NOFS} compared to \ac{SC-OFDM} due to dimension reduction with \ac{NOFS} and \ac{INOFS} transforms. Neural networks are used for these transforms, with tunable connections allowing for pruning of unimportant connections, resulting in lower computational complexity, particularly in multiplication operations, compared to SC-OFDM. } } 

\item{{The waveform framework utilizes shallow neural networks to train optimal signal sub-carriers, ensuring rapid training and providing explainability. The trained model maintains robustness in wireless channels, eliminating the need for re-training with changing channel conditions. This approach is suitable for integrated \ac{AI} and communication in 6G usage scenarios \cite{6G_IMT2030} defined by the International Telecommunication Union (ITU). } }

\item{{A proof-of-concept hardware implementation validates the proposed signal waveform regarding \ac{BER}, \ac{EVM}, spectrum, and modulation constellation performance. The experiment showcases two practical scenarios. The first demonstrates data transmission using fewer spectral resources, while the second shows increased data transmission rate in a given time within the same spectral resources. }} 

\end{itemize}

The rest of the paper is organized as follows. Section \ref{sec:irSinc} briefly describes the principle of irregular Sinc (irSinc). The principle of the multi-carrier \ac{NOFS} is explained in Section \ref{sec:MC-NOFS}, in which system fundamentals and challenges are described. Section \ref{sec:SC_NOFS} describes the proposed \ac{SC-NOFS} signal and shows the neural network training methodology to realize the signal framework. Simulation results on \ac{BER}, \ac{PAPR}, channel impacts, and complexity analysis are presented in Section \ref{sec:performance}. To validate the proposed signal format, hardware implementation is evaluated in Section \ref{sec:hardware_prototyping}. Finally, Section \ref{sec:conclusion} concludes the work.

Notations: Unless otherwise specified, matrices are denoted by bold uppercase letters (i.e., $\mathbf{F}$), vectors are represented by bold lowercase letters (i.e., $\mathbf{x}$, $\mathbf{s}$), and scalars are denoted by normal font (i.e., $\rho$). Subscripts indicate the location of the entry in the matrices or vectors (i.e., $c_{i,j}$ and $s_n$ are the ($i,j$)-th and the $n$-th element in $\mathbf{C}$ and $\mathbf{s}$, respectively)

\section{Irregular Sinc (irSinc)}\label{sec:irSinc}

In practical OFDM communication systems, each subcarrier is shaped by a Sinc as illustrated in Fig. \ref{Fig:shape_subcarrier_NOFS_Sinc_RRC_irSinc_pure}(a). An ideal Sinc has an infinite length and is not implementable in practical systems. Therefore the commonly adopted solution is to
truncate the frequency domain signal via bandlimited filters that are chosen carefully to mitigate the power of sidelobes in the time domain. 
Consequently, this can result in minor out-of-band leakage as demonstrated by the corresponding time pulse in Fig. \ref{Fig:shape_subcarrier_NOFS_Sinc_RRC_irSinc_pure}(b). 

Commonly used non-Sinc signals are shaped in frequency-domain by the \ac{RC} response. The \ac{RC} with a roll-of-factor of 0.3 has a smooth sidelobe as shown in Fig. \ref{Fig:shape_subcarrier_NOFS_Sinc_RRC_irSinc_pure}(c),
where we shall notice that the frequency truncation has negligible effect on the corresponding time domain signal as shown in Fig. \ref{Fig:shape_subcarrier_NOFS_Sinc_RRC_irSinc_pure}(d). 
Specifically, the time domain pulse in Fig. \ref{Fig:shape_subcarrier_NOFS_Sinc_RRC_irSinc_pure}(d) has zero out-of-band leakage but at the cost of a wider time occupancy resulting in potential \ac{ISI}.

Both Sinc and non-Sinc shaped sub-carriers are `Symmetric' while in non-orthogonal signals the symmetric feature might not be the optimal considering \ac{ICI} when the transmission rate is higher than the Nyquist rate~\cite{Tongyang_CSNDSP2016}. Therefore, instead of optimizing traditional `Symmetric' frequency response in Fig. \ref{Fig:shape_subcarrier_NOFS_Sinc_RRC_irSinc_pure}(a)(c), we proposed in \cite{Tongyang_Nature_2023} an `Asymmetric' frequency response pattern, termed as irregular Sinc (irSinc). One example of the irSinc is illustrated in Fig. \ref{Fig:shape_subcarrier_NOFS_Sinc_RRC_irSinc_pure}(e) where its irregular shaping can adaptively mitigate multicarrier \ac{ICI}. Its time domain pulse is illustrated in Fig. \ref{Fig:shape_subcarrier_NOFS_Sinc_RRC_irSinc_pure}(f) with a rectangular shape but with the rippled features at the top. It should be noted that the irSinc in Fig. \ref{Fig:shape_subcarrier_NOFS_Sinc_RRC_irSinc_pure}(e) is just one example. For multicarrier transmissions, each sub-carrier should utilize a unique irSinc signal as illustrated in Fig. \ref{Fig:NOFS_irSinc_subcarrier} in order to improve the communication performance. The shape of each sub-carrier in NOFS signals varies, and it is adjustable based on the signal's specific characteristics. To optimize error performance or in other words to maximize the signal-to-interference power ratio, the irregular Sinc waveform must be tailored for each particular signal configuration. As the number of sub-carriers and sub-carrier spacing compression change, the signal interference characteristics also change accordingly. Thus, it becomes necessary to re-optimize the shape of all irregular Sinc for each sub-carrier to maximize error performance.

\section{Multi-Carrier NOFS} \label{sec:MC-NOFS}

\begin{figure}[t!]
\begin{center}
\includegraphics[scale=0.53]{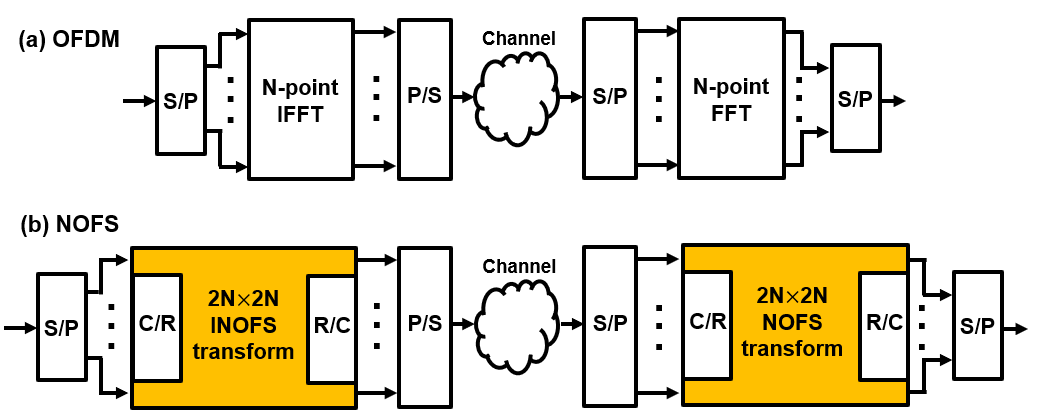}
\end{center}
\caption{Block diagram of (a) multi-carrier OFDM communication link with complex-value signal processing, (b) multi-carrier NOFS communication link with real-value signal processing. C/R indicates complex to real conversion. R/C indicates real to complex conversion.}
\label{Fig:NOFS_OFDM_block_diagram}
\end{figure}

\subsection{System Fundamentals}

\begin{figure}[t!]
\begin{center}
\includegraphics[scale=0.45]{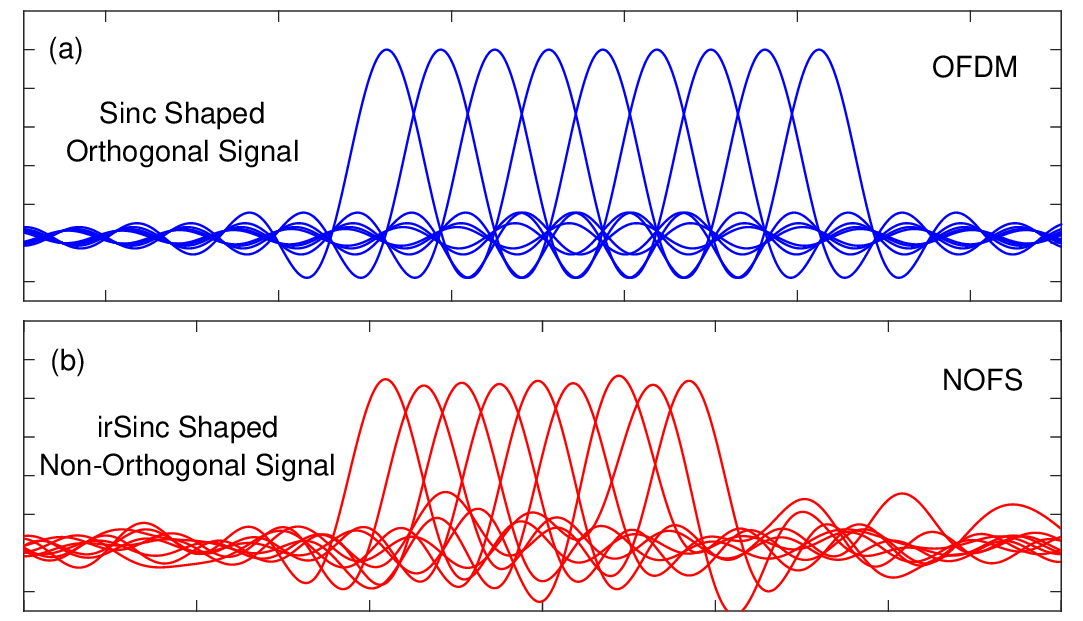}
\end{center}
\caption{Sub-carrier shaping and packing schemes for Sinc shaped OFDM signal and irSinc shaped NOFS signal. Both signals have the same sub-carrier bandwidth while the NOFS signal has compressed sub-carrier spacing.}
\label{Fig:NOFS_irSinc_subcarrier_4}
\end{figure}

In a traditional OFDM system, \ac{IFFT} \cite{Fourier_fast_1965} and \ac{FFT} are used for signal modulation and demodulation in Fig. \ref{Fig:NOFS_OFDM_block_diagram}(a). To realize an irSinc enabled multicarrier system, a new pair of signal transforms, i.e., \ac{INOFS} and \ac{NOFS}, has to be applied as shown in Fig. \ref{Fig:NOFS_OFDM_block_diagram}(b). To show the effect of the irSinc transforms, we demonstrate the \ac{NOFS} signal with the sub-carrier packing scheme in Fig. \ref{Fig:NOFS_irSinc_subcarrier_4}(b). Compared to the OFDM sub-carrier packing in Fig. \ref{Fig:NOFS_irSinc_subcarrier_4}(a), the spectral efficiency of NOFS is improved via non-orthogonally packing sub-carriers closer, where each sub-carrier is shaped by an independent and unique irSinc.

Unlike the complex-value signal processing in Fig. \ref{Fig:NOFS_OFDM_block_diagram}(a), the INOFS and NOFS transforms require real-value signal processing in order to be compatible with neural network architectures. Therefore, considering a complex signal with $N$ sub-carriers, its real-value signal generation model, with a dimension of $2N$, is defined in a neural network format as
\begin{eqnarray}\label{eq:NOFS_Tx}
\mathbf{x} = \sigma(\mathbf{G_\gamma}\mathbf{s}+\Psi_{_{tx}}),
\end{eqnarray}
where the vector $\mathbf{x}\in\mathbb{R}^{2N{\times}1}$ represents the real-value NOFS signal. The matrix $\mathbf{G_\gamma}\in\mathbb{R}^{2N{\times}2N}$ represents the INOFS transform that is similar to the function of \ac{IFFT} in \ac{OFDM}. The subscript $\gamma$ indicates the matrix is in a neural network format. The vector $\mathbf{s}\in\mathbb{R}^{2N{\times}1}$ represents the real-value modulation \ac{QAM} symbols by packing real and imaginary parts together. The vector $\Psi_{_{tx}}\in\mathbb{R}^{2N{\times}1}$ is the transmitter side bias that is used to fine tune the output from the multiplication of $\mathbf{G_\gamma}\mathbf{s}$, and $\sigma(\cdotp)$ is the activation function used to convert the input into a target output.

Before passing through a channel, the transmitter side NOFS signal $\mathbf{x}\in\mathbb{R}^{2N{\times}1}$ is reconstructed into a complex signal as
\begin{eqnarray}\label{eq:NOFS_Tx_complex}
\mathbf{\bar{x}} = \mathbf{x}_{(1:N)}+i\cdotp\mathbf{x}_{(N+1:2N)}. 
\end{eqnarray}

\begin{figure}[t!]
\begin{center}
\includegraphics[scale=0.65]{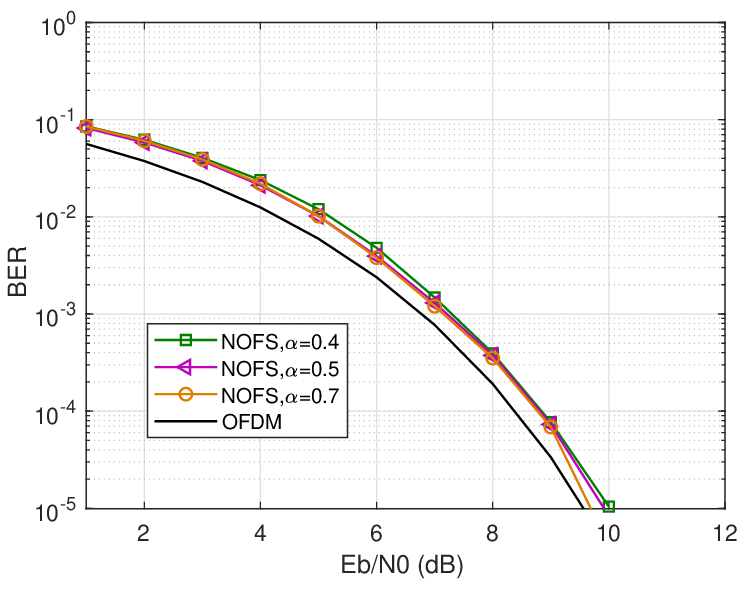}
\end{center}
\caption{BER for multi-carrier NOFS signals. $\alpha$=0.7,0.5,0.4 indicate 43\%,100\%,150\% spectral efficiency improvement, respectively.}
\label{Fig:T_NOFS_4QAM_BER_truncated}
\end{figure}

The complex signal $\mathbf{\bar{x}}\in\mathbb{C}^{N{\times}1}$ passes through a multipath channel $\mathbf{H}\in\mathbb{C}^{N{\times}N}$ and \ac{AWGN} $\mathbf{z}\in\mathbb{C}^{N{\times}1}$ leading to the received signal $\mathbf{\bar{y}}\in\mathbb{C}^{N{\times}1}$ given by
\begin{eqnarray}\label{eq:NOFS_Tx_channel}
\mathbf{\bar{y}} = \mathbf{H}\mathbf{\bar{x}}+\mathbf{z} = \mathbf{H}(\mathbf{x}_{(1:N)}+i\cdotp\mathbf{x}_{(N+1:2N)})+\mathbf{z}.
\end{eqnarray}

At the receiver, assuming $\mathbf{H}$ is well structured, the received signal $\mathbf{\bar{y}}\in\mathbb{C}^{N{\times}1}$ is equalized to $\mathbf{\hat{y}}\in\mathbb{C}^{N{\times}1}$ in time-domain by removing the effect of $\mathbf{H}\in\mathbb{C}^{N{\times}N}$ as
\begin{eqnarray}\label{eq:NOFS_Rx_channel_equalized}
\mathbf{\hat{y}} = \mathbf{H}^{-1}\mathbf{\bar{y}} = \mathbf{x}_{(1:N)}+i\cdotp\mathbf{x}_{(N+1:2N)}+\mathbf{H}^{-1}\mathbf{z}.
\end{eqnarray}

The complex signal $\mathbf{\hat{y}}\in\mathbb{C}^{N{\times}1}$ is converted back to its real value format $\mathbf{{y}}\in\mathbb{R}^{2N{\times}1}$ as
\begin{eqnarray}\label{eq:NOFS_Rx_channel_equalized_real}
\mathbf{{y}}=\mathbf{x}+[\Re(\mathbf{H}^{-1}\mathbf{z});\Im(\mathbf{H}^{-1}\mathbf{z})].
\end{eqnarray}

Then a matched operation of \eqref{eq:NOFS_Tx}, aiming to demultiplex the signals to baseband \ac{QAM} symbols, is given by

\begin{eqnarray}\label{eq:NOFS_Rx_demultiplexed}
\begin{split}
\mathbf{\hat{s}} &= \sigma(\mathbf{G^{*}_\gamma}\mathbf{{y}}+\Psi_{_{rx}})\\
&= \sigma(\mathbf{G^{*}_\gamma}(\sigma(\mathbf{G_\gamma}\mathbf{s}+\Psi_{_{tx}})+\mathbf{\epsilon})+\Psi_{_{rx}}),
\end{split}
\end{eqnarray}
where the term $\mathbf{\epsilon}\in\mathbb{R}^{2N{\times}1}=[\Re(\mathbf{H}^{-1}\mathbf{z});\Im(\mathbf{H}^{-1}\mathbf{z})]$, and $\Psi_{_{rx}}\in\mathbb{R}^{2N{\times}1}$ indicates the receiver side bias.

Utilizing the adaptive irSinc shaping for each sub-carrier, we examine the \ac{BER} performance of the multi-carrier NOFS system, as illustrated in Fig. \ref{Fig:T_NOFS_4QAM_BER_truncated} (reused from \cite{Tongyang_Nature_2023}). The results indicate that, with the bandwidth compression factor $\alpha$=0.4, the NOFS system achieves a significant 60\% saving of communication resources, leading to an impressive 150\% enhancement in spectral efficiency. Notably, it approaches the performance of the traditional OFDM system.

\subsection{Challenges}

The existing 5G standard employs the \ac{OFDM} waveform format where sub-carriers are orthogonally packed leading to simple signal generation, signal detection, and channel estimation. The non-orthogonal multi-carrier NOFS achieves significant spectral efficiency improvement, which is potential to enable new capabilities in future communications. However, regardless of advantageous features, multi-carrier NOFS poses challenges, especially the increased power leakage, high complexity channel estimation/equalization, and incompatibility with existing standards.

\subsubsection{Power Leakage}

{Out-of-band power leakage refers to the emission of signal power outside of the designated frequency band of a communication system. In co-existence communication scenarios, where different signal bands are positioned adjacent to each other in the frequency domain, out-of-band power leakage can have significant implications. It can lead to interference with neighbouring communication systems operating in adjacent frequency bands, thereby affecting their performance and reliability.} For the bandwidth compressed \ac{NOFS}, its sub-carrier packing pattern is illustrated in Fig. \ref{Fig:NOFS_irSinc_subcarrier_4} with irregular out-of-band patterns. Therefore, it is obvious that the NOFS signal waveform may introduce higher out-of-band power leakage than other waveforms. Such a high out-of-band power leakage signal design is not suitable for co-existence communication scenarios because the leakage would cause interference to adjacent signals. Some non-orthogonal signals cope with this issue relying on pulse shaping at the cost of increased signal processing complexity. Therefore, aiming for wider application scenarios, optimization of the out-of-band power is of importance.

\begin{figure*}[t!]
\begin{center}
\includegraphics[scale=0.55]{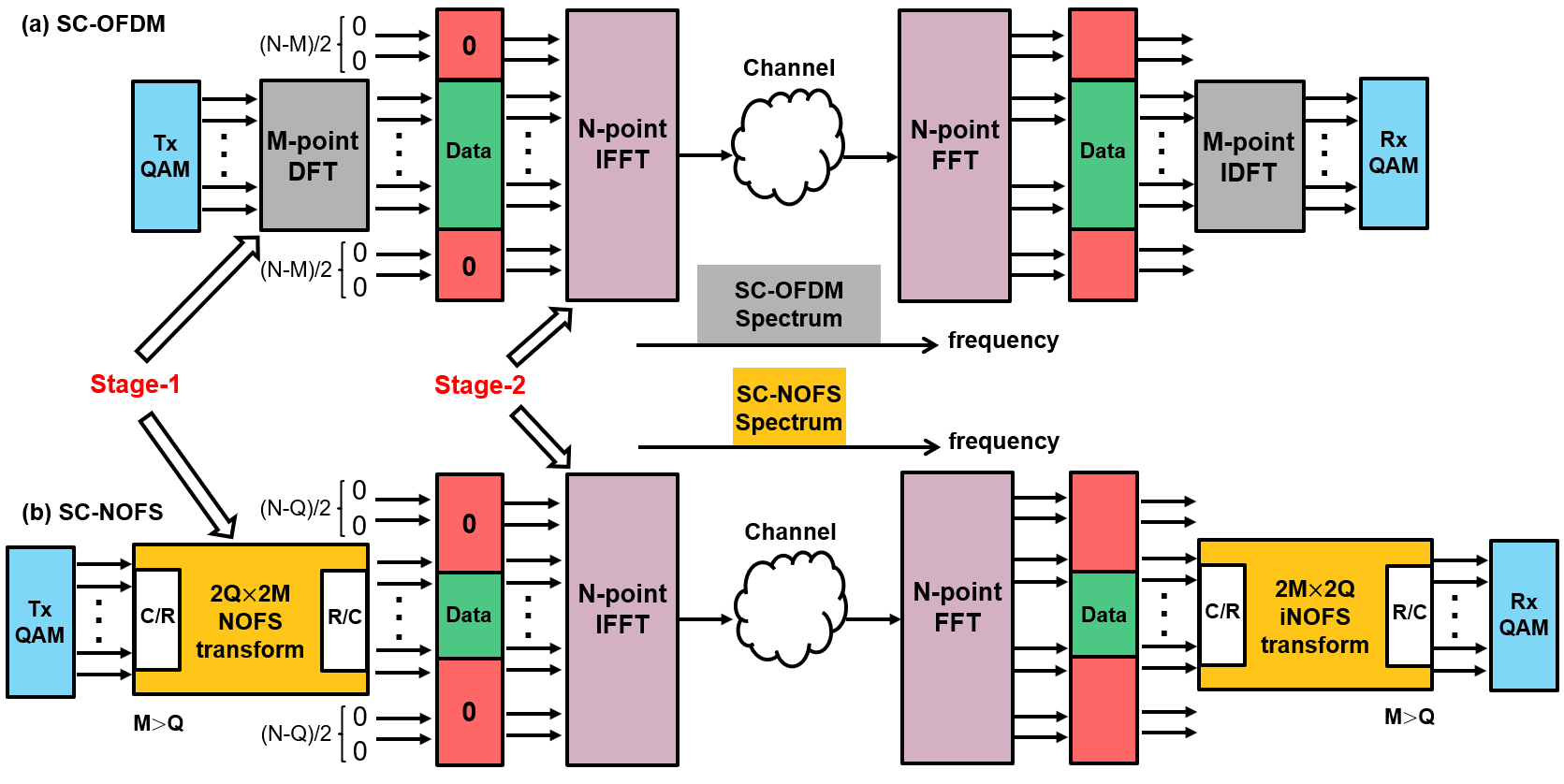}
\end{center}
\caption{Block diagram of (a) single-carrier OFDM communication link with complex-value signal processing, (b) single-carrier NOFS communication link with real-value signal processing. C/R indicates complex to real conversion. R/C indicates real to complex conversion. }
\label{Fig:SC_NOFS_system_block_diagram}
\end{figure*}

\subsubsection{Channel Estimation and Equalization}

Channel estimation and equalization for non-orthogonal signals are very challenging due to the self-created \ac{ICI} such that the commonly used frequency-domain one-tap channel estimation and equalization are no longer applicable. In NOFS, due to the irSinc frequency response on each sub-carrier, the conventional Fourier transform is not applicable. Therefore, time-domain channel estimation and equalization may perform better than  frequency domain counterparts. However, time-domain operations require de-convolution processing, which is more complicated than the frequency-domain approach and maybe impractical in real world communication systems. 

For ease of explanation, we consider multipath frequency selective channel models and ignore Doppler effects. To estimate the channel from the distorted NOFS signal in \eqref{eq:NOFS_Tx_channel}, pilot symbols are needed prior to data symbols. For each pilot symbol, \ac{CP} is added to mitigate \ac{ISI} and simplify channel estimation. The mathematical model of pilot symbols going through the channel is given by
\begin{eqnarray}\label{eq:NOFS_multipath_pilot_CP}
\mathbf{\bar{y}_{_{cp}}} = \mathbf{H}\mathbf{\bar{x}_{_{cp}}}+\mathbf{z_{_{cp}}},
\end{eqnarray}
where $\mathbf{\bar{x}_{_{cp}}}\in\mathbb{C}^{N_{cp}{\times}1}$ is the transmitted pilot symbol including both \ac{CP} and pilot data. $N_{cp}=N+L$ where $L$ indicates the length of \ac{CP}. The channel matrix $\mathbf{H}\in\mathbb{C}^{N_{cp}{\times}N_{cp}}$ becomes larger due to the addition of \ac{CP}, and $\mathbf{z_{_{cp}}}\in\mathbb{C}^{N_{cp}{\times}1}$ indicates the \ac{AWGN} vector. At the receiver, after removing \ac{CP}, the Toeplitz channel matrix $\mathbf{H}\in\mathbb{C}^{N_{cp}{\times}N_{cp}}$ becomes a circulant channel matrix $\mathbf{H_c}\in\mathbb{C}^{N{\times}N}$ with the following new format

\begin{eqnarray}\label{eq:NOFS_multipath_pilot_remove_CP}
\mathbf{\bar{y}_p} &= \mathbf{H_c}\mathbf{\bar{x}_p}+\mathbf{z},
\end{eqnarray}
where $\mathbf{\bar{x}_p}\in\mathbb{C}^{N{\times}1}$ is the transmitted pilot vector, $\mathbf{\bar{y}_p}\in\mathbb{C}^{N{\times}1}$ is the received and distorted pilot vector.

The relationship between a real-value matrix and a complex-value matrix is given in the following
\begin{equation}\label{eq:NOFS_real_to_complex_matrix_1}
\mathbf{G_\gamma} =
 \begin{bmatrix}
  \Re(\mathbf{\bar{G}_\gamma}) & -\Im(\mathbf{\bar{G}_\gamma})  \\
  \Im(\mathbf{\bar{G}_\gamma}) & \Re(\mathbf{\bar{G}_\gamma})  \\
 \end{bmatrix}
\in\mathbb{R}^{2N{\times}2N},
\end{equation}
where $\mathbf{\bar{G}_\gamma}\in\mathbb{C}^{N{\times}N}$ is the complex-value neural network based NOFS generation matrix with the following format

\begin{equation}\label{eq:NOFS_real_to_complex_matrix_2}
\mathbf{\bar{G}_\gamma} =
 \begin{bmatrix}
  \mathbf{g}_1  \\
  \mathbf{g}_2  \\
  \vdots        \\
  \mathbf{g}_N  \\
 \end{bmatrix}
\in\mathbb{C}^{N{\times}N},
\end{equation}
where $\mathbf{g}_i$ indicates the $i^{th}$ row in $\mathbf{\bar{G}_\gamma}$. To simplify the analysis, considering $\mathbf{\bar{x}_p}$=$\mathbf{x_p}_{(1:N)}+1i\cdotp\mathbf{x_p}_{(N+1:2N)}$ and $\mathbf{x_p}$=$\sigma(\mathbf{G_\gamma}\mathbf{s_p}+{\Psi_{_{tx}}})$, \eqref{eq:NOFS_multipath_pilot_remove_CP} is converted to

\begin{eqnarray}\label{eq:NOFS_multipath_pilot_remove_CP3}
\mathbf{\bar{y}_p}=\mathbf{H_c}(\sigma_c(\mathbf{\bar{G}_\gamma}\mathbf{\bar{s}_p}+{\bar\Psi_{_{tx}}}))+\mathbf{z},
\end{eqnarray}
where $\mathbf{\bar{s}_p}\in\mathbb{C}^{N{\times}1}$ represents complex-value QAM symbols, ${\bar\Psi_{_{tx}}}\in\mathbb{C}^{N{\times}1}$ indicates the transmitter side bias in a complex format. In practice, the signal processing inside an activation function should be in real-value formats due to the characteristics of neural network operations. However, to simplify the expression here, we use a unique activation function $\sigma_c(\cdotp)$ to indicate that the values in the bracket could be complex. But in practical operations, the computation is still in real-value formats. Due to the circulant channel matrix feature, $\mathbf{{H}_c}$ can be reconstructed via copying and shifting $\mathbf{h}\in\mathbb{C}^{N{\times}1}$ at each column. Therefore, \eqref{eq:NOFS_multipath_pilot_remove_CP3} is further converted to
\begin{equation}\label{eq:NOFS_multipath_pilot_rearrange}
\mathbf{\bar{y}_p}=\sigma_c(\mathbf{P})\mathbf{h}+\mathbf{z},
\end{equation}
where $\mathbf{P}=(\mathbf{P}_1+\mathbf{P}_2)\in\mathbb{C}^{N{\times}N}$ is a circulant matrix defined by

\begin{equation}\label{eq:p1_matrix}
\mathbf{P}_1 =
 \begin{bmatrix}
  \mathbf{g}_{1} & \mathbf{g}_{Q}     & \cdots &   \mathbf{g}_2 \\
  \mathbf{g}_{2} & \mathbf{g}_{1} & \ddots &       \vdots \\
  \vdots  & \ddots  & \ddots &   \mathbf{g}_{Q}\\
  \mathbf{g}_{Q} & \cdots & \mathbf{g}_2 & \mathbf{g}_1
 \end{bmatrix}\cdotp
 \begin{bmatrix}
  \mathbf{\bar{s}_p} & 0 & \cdots & 0 \\
  0 & \mathbf{\bar{s}_p} & \cdots & 0 \\
  \vdots  & \vdots  & \ddots & \vdots  \\
  0 & 0 & \cdots & \mathbf{\bar{s}_p}
 \end{bmatrix}
\end{equation}

\begin{equation}\label{eq:p2_matrix}
\mathbf{P}_2 =
 \begin{bmatrix}
  \bar\Psi_{_{tx,1}} & \bar\Psi_{_{tx,Q}}     & \cdots &   \bar\Psi_{_{tx,2}} \\
  \bar\Psi_{_{tx,2}} & \bar\Psi_{_{tx,1}} & \ddots &       \vdots \\
  \vdots  & \ddots  & \ddots &   \bar\Psi_{_{tx,Q}}\\
  \bar\Psi_{_{tx,Q}} & \cdots & \bar\Psi_{_{tx,2}} & \bar\Psi_{_{tx,1}}
 \end{bmatrix}
\end{equation}
where $\bar\Psi_{_{tx,i}}$ indicates the $i^{th}$ element in $\bar\Psi_{_{tx}}$. The channel elements are therefore computed as
\begin{equation}\label{eq:h_estimate}
\mathbf{\hat{h}}=\mathbf{\sigma_c(P)^{*}(\sigma_c(P)\sigma_c(P)^{*})^{-1}}\mathbf{\bar{y}_p}.
\end{equation}

It is apparent from the above signal processing that time-domain channel estimation is far more complicated than its frequency-domain operations because matrix multiplications and matrix inverse calculations are needed for the time-domain computations. In addition, channel equalization relies on the reconstruction of the channel matrix resulting in extra matrix multiplications and matrix inverse operations. Hence, there is a pressing demand for a novel signal design that maintains compatibility with frequency-domain channel estimation and equalization operations.

\subsubsection{Compatibility}

Compatibility with available standards is a big challenge for all advanced physical layer signal waveform design. Operators have invested huge \cite{NGMN_6G_statement, 6G_industry_need_2023} on hardware infrastructure where OFDM is integrated as the physical layer waveform. A fundamentally new physical layer design will require replacement of existing hardware, which is not a wise investment when previous investment has not been repaid. In available standards, sub-carriers are shaped by conventional Sinc where it can be implemented directly by using \ac{IFFT}. For NOFS, its irregular Sinc shaped sub-carriers change the physical layer characteristics fundamentally and will require new hardware to replace the OFDM supported hardware. In addition, the irregular Sinc sub-carriers would also cause interference to already deployed standards. Therefore, there is a critical need for a physical layer signal design that inherits the advantages of NOFS signals while ensuring compatibility with existing standards.

\section{Single-Carrier NOFS}\label{sec:SC_NOFS}

\subsection{Proposed SC-NOFS Signal Framework}\label{subsec:SC-NOFS}

Due to the simple signal processing and robustness to multipath channels, \ac{OFDM} has been standardized in 4G, 5G, WiFi, and other relevant systems. Therefore, to enable straightforward integration in existing standards, new physical layer waveforms could adopt advantages from OFDM through a two-stage waveform generation framework.

A conventional two-stage signal framework for SC-OFDM is illustrated in Fig. \ref{Fig:SC_NOFS_system_block_diagram}(a) where the first stage is an $M$-point \ac{DFT} \cite{Fourier_transform_1822} on an $M$-point \ac{QAM} symbol sequence and the second stage follows the traditional $N$-point OFDM signal generation. However, the original motivation of SC-OFDM is to reduce \ac{PAPR} but without any gains in spectral efficiency. 

Therefore, we propose and demonstrate a two-stage SC-NOFS signal framework in Fig. \ref{Fig:SC_NOFS_system_block_diagram}(b) where the user-defined signal generation is integrated at the first stage and OFDM signal generation is applied at the second stage such that over-the-air signal format still follows the existing OFDM standard. To enhance spectral efficiency, we intentionally use a $2Q\times{2M}$ NOFS transform such that the $2Q$-point output is shorter than its $2M$-point input leading to a compression factor defined as $\alpha$=$Q/M$. Therefore, a reduced number of data sub-carriers are required for the second stage, leading to a reduction in occupied spectral resources, as illustrated in Fig. \ref{Fig:SC_NOFS_system_block_diagram}, ultimately leading to improved spectral efficiency. It is noted that C/R and R/C modules are required to convert between complex-value symbols and real-value symbols because the NOFS and INOFS transforms are operated based on real-value neural network architectures. 

The benefits of the two-stage SC-NOFS waveform processing are

\begin{itemize} 

\item{ SC-NOFS achieves higher spectral efficiency than SC-OFDM because a spectrally efficient waveform pattern is integrated at the first stage. } 
\item{ It is compatible with existing standards since over-the-air interface is in the OFDM format.  } 
\item{ Its out-of-band power leakage is reduced and is within the standard requirement. }
\item{ Its \ac{PAPR} is reduced compared to its multi-carrier version. }
\item{ Channel estimation and equalization follow the traditional frequency-domain one-tap signal processing. }
\end{itemize}

\begin{figure*}[t!]
\begin{center}
\includegraphics[scale=0.58]{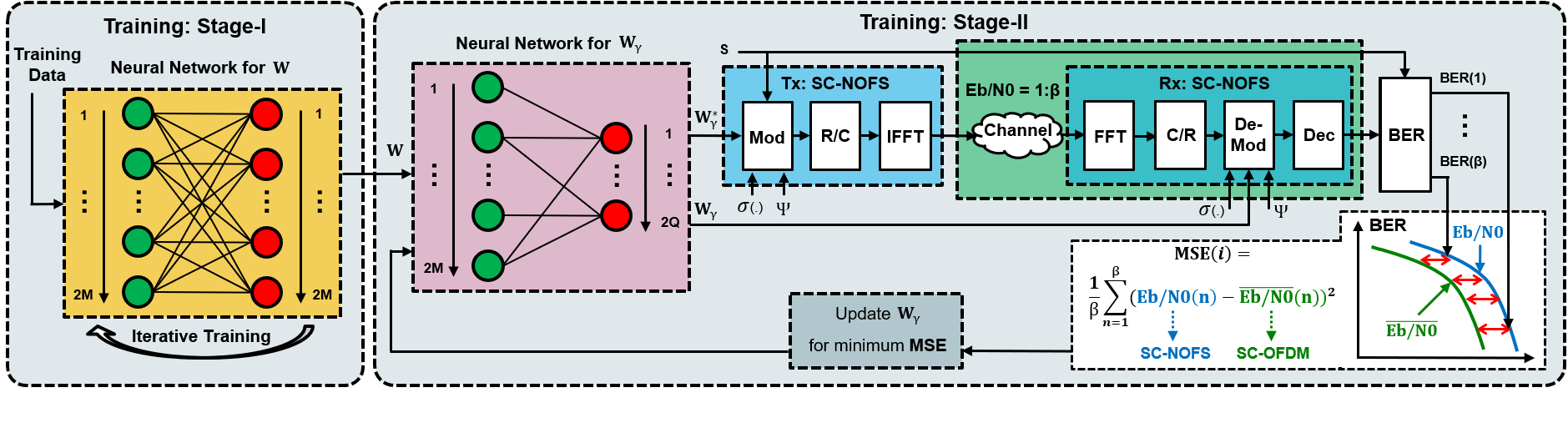}
\end{center}
\caption{Training mechanism for SC-NOFS.  }
\label{Fig:SC_NOFS_training_block_diagram}
\end{figure*}

In the first stage, the $2Q\times{2M}$ real-value NOFS transform aims to realize the following
\begin{eqnarray}\label{eq:SC_NOFS_Tx_stage_1}
\mathbf{x_a} = \sigma(\mathbf{W^{*}_\gamma}\mathbf{s}+\Psi_{_{tx}}),
\end{eqnarray}
where the matrix $\mathbf{W^{*}_\gamma}\in\mathbb{R}^{2Q{\times}2M}$ indicates the first-stage neural network based NOFS transform where $M>Q$. The aim of the non-square neural network matrix $\mathbf{W^{*}_\gamma}$ is to output a shorter symbol vector than its input symbol vector for the purpose of spectral efficiency enhancement. The vector $\mathbf{s}\in\mathbb{R}^{2M{\times}1}$ represents the real-value \ac{QAM} symbols. Due to the dimension reduction of $\mathbf{W^{*}_\gamma}$, the bias $\Psi_{_{tx}}\in\mathbb{R}^{2Q{\times}1}$ and the output $\mathbf{x_a}\in\mathbb{R}^{2Q{\times}1}$ become shorter.

The complex signal $\mathbf{\bar{x}_a}\in\mathbb{C}^{Q{\times}1}$ is reconstructed from $\mathbf{x_a}\in\mathbb{R}^{2Q{\times}1}$ and is symmetrically zero padded to obtain $\mathbf{\bar{\bar{x}}_a}\in\mathbb{C}^{N{\times}1}$ for the second stage $N$-point \ac{IFFT} processing as

\begin{eqnarray}\label{eq:SC_NOFS_Tx_stage_2}
\mathbf{\bar{x}_b} = \mathbf{F}\mathbf{\bar{\bar{x}}_a},
\end{eqnarray}
where the matrix $\mathbf{F}\in\mathbb{C}^{N{\times}N}$ represents the \ac{IFFT} matrix.

The second stage signal $\mathbf{\bar{x}_b}$ goes through the multipath channel $\mathbf{H}\in\mathbb{C}^{N{\times}N}$ and \ac{AWGN} $\mathbf{z}\in\mathbb{C}^{N{\times}1}$ giving an expression as
\begin{eqnarray}\label{eq:SC_NOFS_Tx_channel}
\mathbf{\bar{y}} = \mathbf{H}\mathbf{\bar{x}_b}+\mathbf{z} = \mathbf{H}\mathbf{F}\mathbf{\bar{\bar{x}}_a}+\mathbf{z}.
\end{eqnarray}

At the receiver, unlike the time-domain channel equalization processing in \eqref{eq:NOFS_Rx_channel_equalized}, the proposed SC-NOFS signal operates \ac{FFT} first to convert the time-domain signal to its frequency-domain version as
\begin{eqnarray}\label{eq:SC_NOFS_Rx_1}
\begin{split}
\mathbf{r} = \mathbf{F^*}\mathbf{\bar{y}} &= \mathbf{F^*}\mathbf{H}\mathbf{F}\mathbf{\bar{\bar{x}}_a}+\mathbf{F^*}\mathbf{z}\\
&=\mathbf{D}\mathbf{\bar{\bar{x}}_a}+\mathbf{F^*}\mathbf{z},
\end{split}
\end{eqnarray}
where $\mathbf{F^*}\in\mathbb{C}^{N{\times}N}$ indicates the \ac{FFT} matrix. The multiplication term $\mathbf{F^*}\mathbf{H}\mathbf{F}$ leads to a diagonal matrix $\mathbf{D}$ which simplifies channel estimation and equalization into one-tap processing. 

The channel frequency response at each sub-carrier, represented by the diagonal elements $diag(\mathbf{D}) = [{d}_1, {d}_2,..., {d}_N]$ in $\mathbf{D}$, is extracted to equalize signals as

\begin{eqnarray}\label{eq:SC_NOFS_Rx_channel_equalized}
\begin{split}
\mathbf{\hat{y}} &= diag(\mathbf{D})^{-1}\mathbf{r} \\
&= \mathbf{\bar{\bar{x}}_a}+diag(\mathbf{D})^{-1}\mathbf{F^*}\mathbf{z} \\
&= \mathbf{\bar{\bar{x}}_a}+ \mathbf{\xi},
\end{split}
\end{eqnarray}
where $\mathbf{\xi}\in\mathbb{C}^{N{\times}1}$. The complex signal $\mathbf{\hat{y}}\in\mathbb{C}^{N{\times}1}$ is truncated to remove the symmetrically padded zeros and is converted back to its real value format $\mathbf{{y}}\in\mathbb{R}^{2Q{\times}1}$ as 

\begin{eqnarray}\label{eq:SC_NOFS_Rx_channel_equalized2}
\begin{split}
\mathbf{{y}} &= \mathbf{{{x}}_a}+ [\Re(\mathbf{\xi}_{((N-Q)/2+1:N/2)});\Im(\mathbf{\xi}_{(N/2+1:N/2+(N-Q)/2)})] \\
&= \sigma(\mathbf{W^{*}_\gamma}\mathbf{s}+\Psi_{_{tx}})+ \mathbf{\varrho},
\end{split}
\end{eqnarray}
where $\mathbf{\varrho}\in\mathbb{R}^{2Q{\times}1}$. Then the INOFS transform on \eqref{eq:SC_NOFS_Rx_channel_equalized2} is given by

\begin{eqnarray}\label{eq:SC_NOFS_Rx_demultiplexed}
\begin{split}
\mathbf{\hat{s}} &= \sigma(\mathbf{W_\gamma}\mathbf{{y}}+\Psi_{_{rx}})\\
&= \sigma(\mathbf{W_\gamma}(\sigma(\mathbf{W^{*}_\gamma}\mathbf{s}+\Psi_{_{tx}})+\mathbf{\varrho})+\Psi_{_{rx}}),
\end{split}
\end{eqnarray}
where $\Psi_{_{rx}}\in\mathbb{R}^{2M{\times}1}$ indicates the receiver side bias.

\subsection{Neural Network Training Framework}

The NOFS/INOFS transform blocks in Fig. \ref{Fig:SC_NOFS_system_block_diagram}(b) take advantages of machine learning to provide new features and achieve better signal performance. The aim of the NOFS transform block is to partially remove some samples such that the total transmission samples are reduced leading to increased spectral efficiency. The removal of samples causes information loss and therefore difficulties in accurate signal recovery. Therefore, the neural network has to be specifically trained to generate optimal output sequence such that original signal can be recovered easily by using its INOFS transform at the receiver.

The training mechanism of the NOFS/INOFS transform neural networks $\mathbf{W^{*}_\gamma}$/$\mathbf{W_\gamma}$ are detailed in Fig. \ref{Fig:SC_NOFS_training_block_diagram}. The first stage training aims to learn legacy signals and emulate them using neural network architectures, which is an efficient mechanism to transform sophisticated signal model into simple neural network architectures with fine-tuning capabilities for each neural connection. In this work, the legacy signal is in an OFDM format but it could be any signals when considering different application scenarios. It is noted that conventional mathematical models have solid theoretical analysis but further improvement or optimization is challenging. Neural networks include an array of internal connections and each connection is associated with a weight to determine the importance of this connection. The obvious advantage of neural networks is its tuning capability where changing internal neural connections can bring new features. This is the reason why in the first stage the framework needs to learn and construct signal models in a neural network format.

The second stage of training aims to fine-tune the neural network obtained from the first stage by optimizing neural connection topology and their weights through sample dimension truncation and neural connection truncation. The objective is to obtain a fine-tuned neural network matrix $\mathbf{W_\gamma}$ that can enhance spectral efficiency and deal with non-orthogonal signal interference. In the second stage, the trained neural network matrix $\mathbf{W^{*}_\gamma}$ is used to modulate $\mathbf{s}$ in the \textit{Mod} block with the manipulations of activation function $\sigma(\cdotp)$ and bias $\Psi$. The SC-NOFS signal is obtained after the \textit{R/C} and \textit{IFFT} blocks, and is distorted by the \ac{AWGN} channel. It is noted that the framework intentionally removes frequency selective channel effects in the training process to avoid variation effects on neural networks training. The channel effect will be removed using traditional channel estimation and equalization algorithms. At the receiver, the signal is demultiplexed at the \textit{FFT} block and is demodulated in the \textit{De-Mod} block using the INOFS transform. At this step, original symbols are recovered but with interference. The optimal signal detection scheme is based on the maximum likelihood principle, which is not the preferred solution for communication sustainability and net-zero objectives. Moreover, the maximum likelihood method is not scalable and its processing complexity is exponentially increased with the signal size. Therefore, a simple signal detection approach, termed \ac{ID} \cite{TongyangCL2013}, based on successive interference cancellation, is preferred in the \textit{Dec} block.

The training objective in Fig. \ref{Fig:SC_NOFS_training_block_diagram} is to achieve the best \ac{BER} performance and therefore minimizing the \ac{BER} gap between the conventional SC-OFDM and the target SC-NOFS is the focus. The gap is calculated based on \ac{MSE}, and once a smaller \ac{MSE} is obtained, the training iteration continues. Otherwise, the training stops with the current neural network matrix $\mathbf{W_\gamma}$ as the final solution. {We use 600,000 4QAM symbols to train the neural network and $10^7$ 4QAM symbols to test the system. The learning rate is 0.01 and the training loss is calculated based on \ac{MSE}. To maintain signal features during training, we use linear activation functions $\sigma(\cdot)$. Random values are initially given to the elements in the neural network and the optimal values will be obtained after 1000 training iterations via back propagation using the stochastic gradient descent (SGD) optimization algorithm.} The training framework is flexible and it can set up other training objectives such as to achieve low \ac{PAPR} or to cut out-of-band power leakage.

\section{Performance Evaluation}\label{sec:performance}

To show the advantages of our proposed SC-NOFS signal framework, we include two related works in \ac{HC-MCM} \cite{HC_MCM_2004} and truncated-OFDM \cite{Truncated_OFDM_2016}, where only one-stage signal generation with very limited number of sub-carriers were investigated. To have fair comparisons, we have to integrate the one-stage signal techniques from \cite{HC_MCM_2004, Truncated_OFDM_2016} into our proposed two-stage single-carrier signal framework using the structure in Fig. \ref{Fig:SC_NOFS_system_block_diagram}. In addition, it is noted that the works in \cite{HC_MCM_2004, Truncated_OFDM_2016} utilize a small number of sub-carriers because they have to rely on maximum likelihood or sphere decoding signal detection algorithms. In contrast, our approach allows for the utilization of 1024 sub-carriers while employing simpler signal detection methods. The key novelty of SC-NOFS is the NOFS/INOFS transform aiming to generate a shorter sequence relying on irSinc shaped sub-carrier and artificial intelligence. Conventional signal truncation solutions in \ac{HC-MCM} \cite{HC_MCM_2004} and truncated-OFDM \cite{Truncated_OFDM_2016} have limitations because their signals are Sinc shaped, which is inflexible and is not the optimal solution for non-orthogonal signals \cite{Tongyang_Nature_2023}. To simplify the notations, we use `Sinc-Traditional(\cite{HC_MCM_2004},\cite{Truncated_OFDM_2016})' to indicate the benchmark in the comparison of the following results.

\subsection{BER Performance}

\begin{figure}[t]
\begin{center}
\includegraphics[scale=0.59]{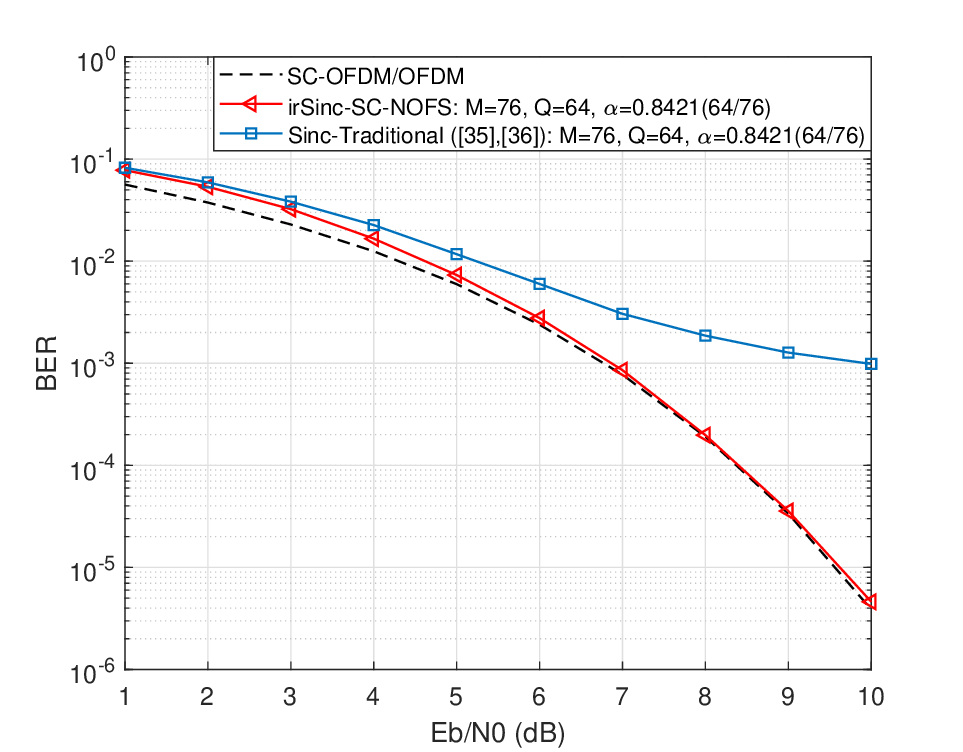}
\end{center}
\caption{BER performance in AWGN for SC-NOFS signals with $Q$=64, $M$=76, $N$=128, $\alpha$=$Q/M$=0.8421, spectral efficiency improvement is $(M/Q-1)\times$100\%=18.75\%. }
\label{Fig:T_NOFS_4QAM_rho_1_alpha_1_N_76}
\end{figure}

\begin{figure}[t]
\begin{center}
\includegraphics[scale=0.59]{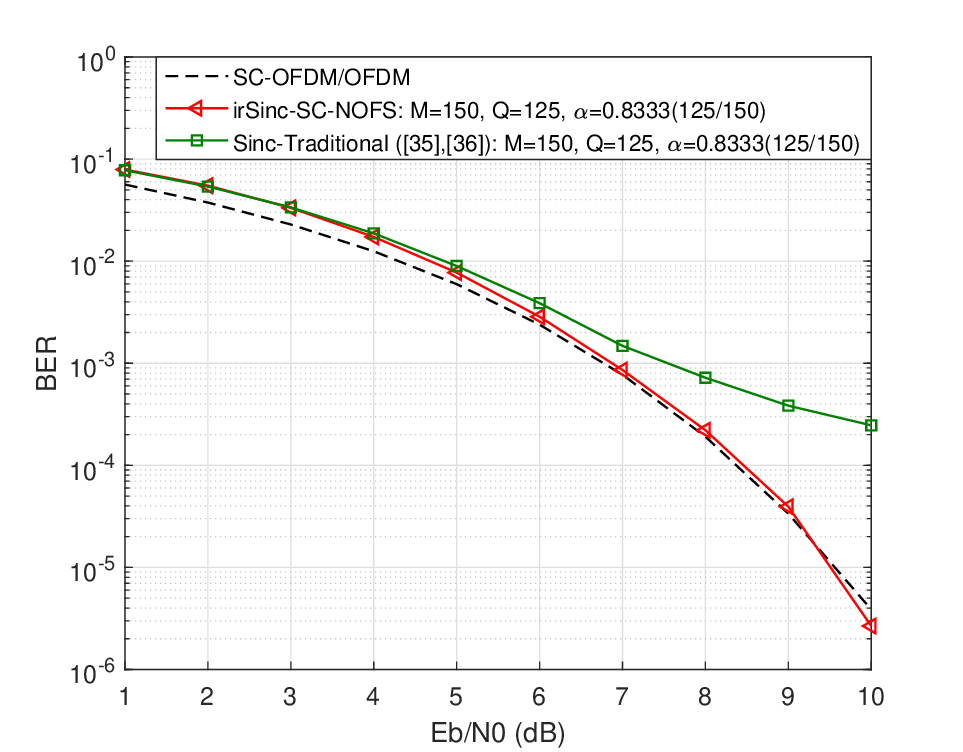}
\end{center}
\caption{BER performance in AWGN for SC-NOFS signals with $Q$=125, $M$=150, $N$=256, $\alpha$=$Q/M$=0.8333, spectral efficiency improvement is $(M/Q-1)\times$100\%=20\%. }
\label{Fig:T_NOFS_4QAM_rho_1_alpha_1_N_150}
\end{figure}

\begin{figure}[t]
\begin{center}
\includegraphics[scale=0.59]{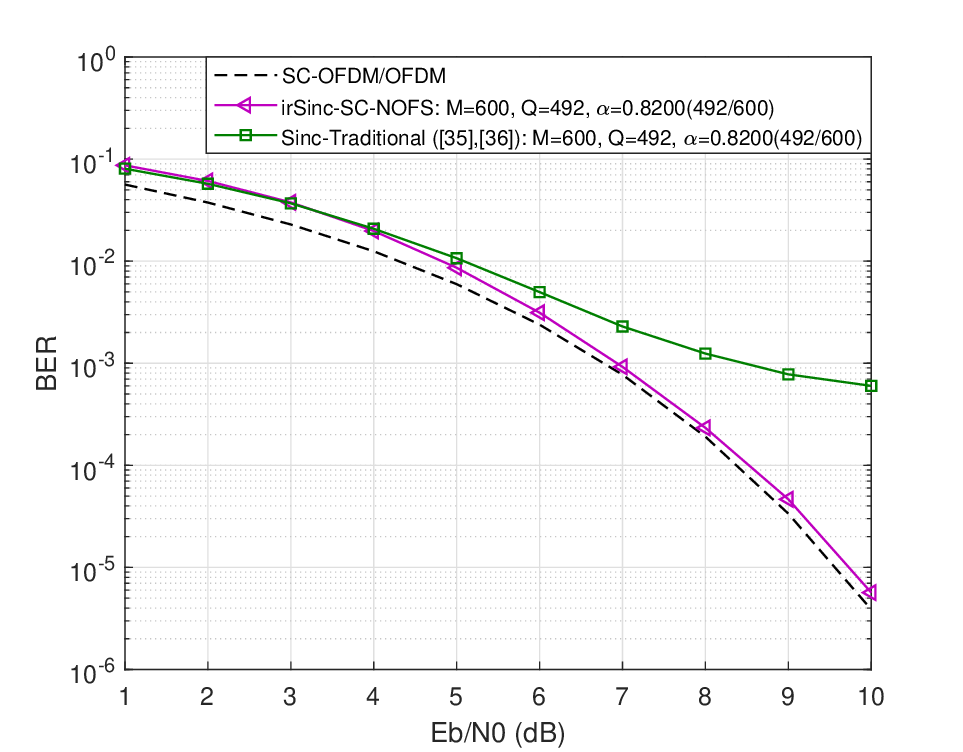}
\end{center}
\caption{BER performance in AWGN for SC-NOFS signals with $Q$=492, $M$=600, $N$=1024, $\alpha$=$Q/M$=0.82, spectral efficiency improvement is $(M/Q-1)\times$100\%=22\%. }
\label{Fig:T_NOFS_4QAM_rho_1_alpha_1_N_600}
\end{figure}

We firstly evaluate the signal performance in Fig. \ref{Fig:T_NOFS_4QAM_rho_1_alpha_1_N_76} with $M$=76 input 4QAM symbols and $N$=128. Based on the training methodology in Fig. \ref{Fig:SC_NOFS_training_block_diagram}, the proposed SC-NOFS signal can reach the compression factor $\alpha$=64/76=0.8421, where 12 output samples are removed out of 76 samples. Its performance curve is overlapping with the SC-OFDM/OFDM indicating the performance reliability and 18.75\%=(76/64-1)$\times$100\% spectral efficiency improvement. On the other hand, under the same compression ratio, the Sinc-Traditional(\cite{HC_MCM_2004},\cite{Truncated_OFDM_2016}) signals are not recoverable and their performance obviously moves away from SC-OFDM/OFDM. 

It is noted that the size of an NOFS/INOFS transform matrix is determined by the number of sub-carriers. The larger size of the NOFS/INOFS transform matrix, the more flexibility of the neural network architecture because more neural connections and weights are available to be tuned. As the number of sub-carriers decreases, the benefits of signal compression gradually diminish owing to the reduced flexibility in neural network tuning. Therefore, it is of great importance to integrate more neural connections in the transform matrix. One efficient solution is to increase the number of input samples $M$ such that the size of the generator matrix is increased and the diversity of matrix elements is enhanced.

In Fig. \ref{Fig:T_NOFS_4QAM_rho_1_alpha_1_N_150}, we increase the number of input samples to $M$=150. Results show clearly that the higher number of input samples enables better spectral efficiency because the compression factor can be further reduced to $\alpha$=0.8333. On the other hand, for those Sinc shaped solutions in \cite{HC_MCM_2004}\cite{Truncated_OFDM_2016}, the performance curve is away from SC-OFDM/OFDM. When we further increase the number of input samples to $M$=600 in Fig. \ref{Fig:T_NOFS_4QAM_rho_1_alpha_1_N_600}, the spectral efficiency is further improved with the compression factor reduced to $\alpha$=0.82.

\subsection{PAPR Performance}

\begin{figure}[t]
\begin{center}
\includegraphics[scale=0.43]{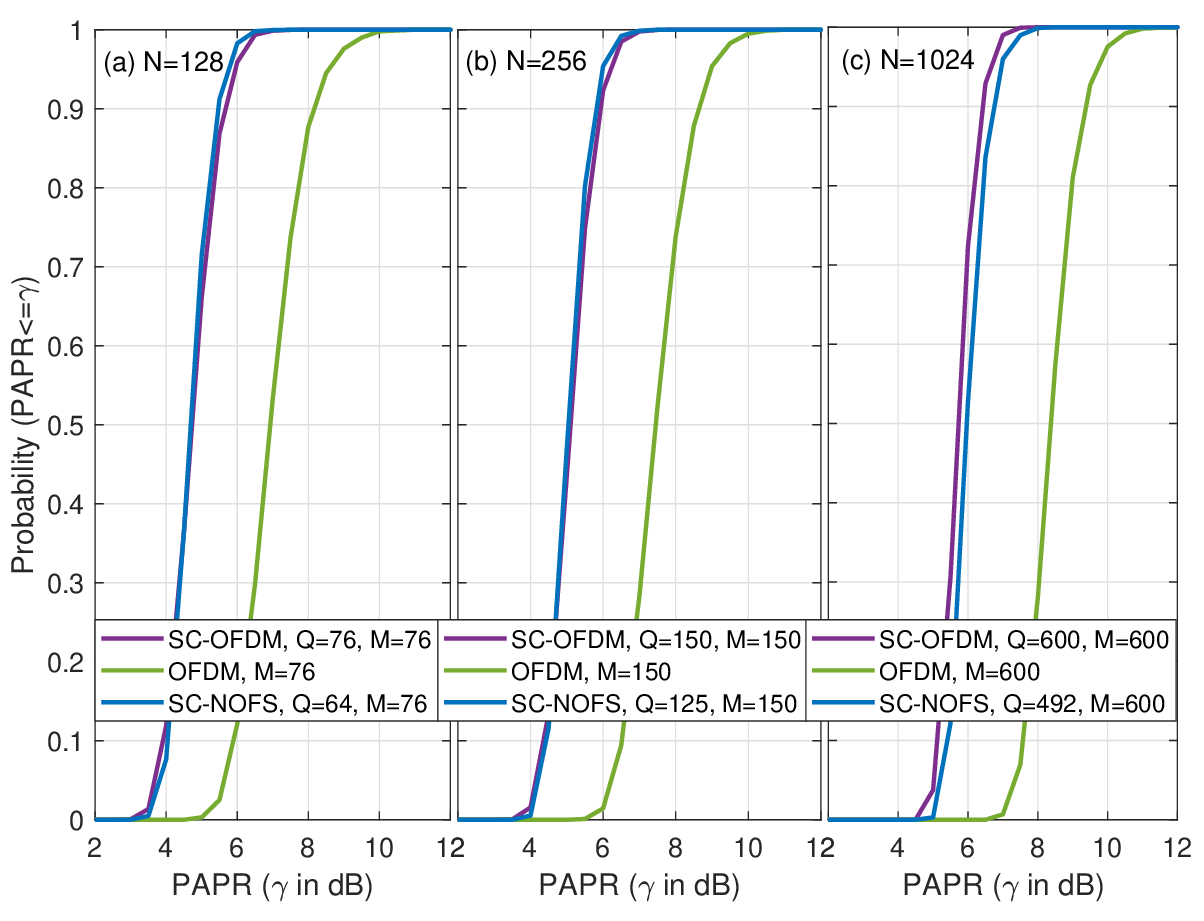}
\end{center}
\caption{PAPR performance for SC-NOFS signals. }
\label{Fig:PAPR_N128_256_1024_SC_NOFS_TOFDM_OFDM}
\end{figure}

In Fig. \ref{Fig:PAPR_N128_256_1024_SC_NOFS_TOFDM_OFDM}, we evaluate the \ac{PAPR} of SC-NOFS signals with different numbers of sub-carriers. Firstly, both single-carrier signals (SC-NOFS and SC-OFDM) exhibit a noticeable improvement in PAPR compared to traditional multi-carrier OFDM signals. This improvement is reasonable since the overlap of multiple sub-carriers in OFDM leads to a high peak power. Secondly, SC-NOFS signals demonstrate nearly identical PAPR when compared to SC-OFDM, with only minor differences observed. It should be emphasized that the primary goal of training or optimizing irSinc in SC-NOFS is to mitigate intercarrier interference, rather than focusing on improving PAPR. Therefore, any improvement in PAPR achieved through the use of irSinc is considered an added benefit.

\subsection{Frequency Selective Channel Performance}

\begin{figure}[t]
\begin{center}
\includegraphics[scale=0.59]{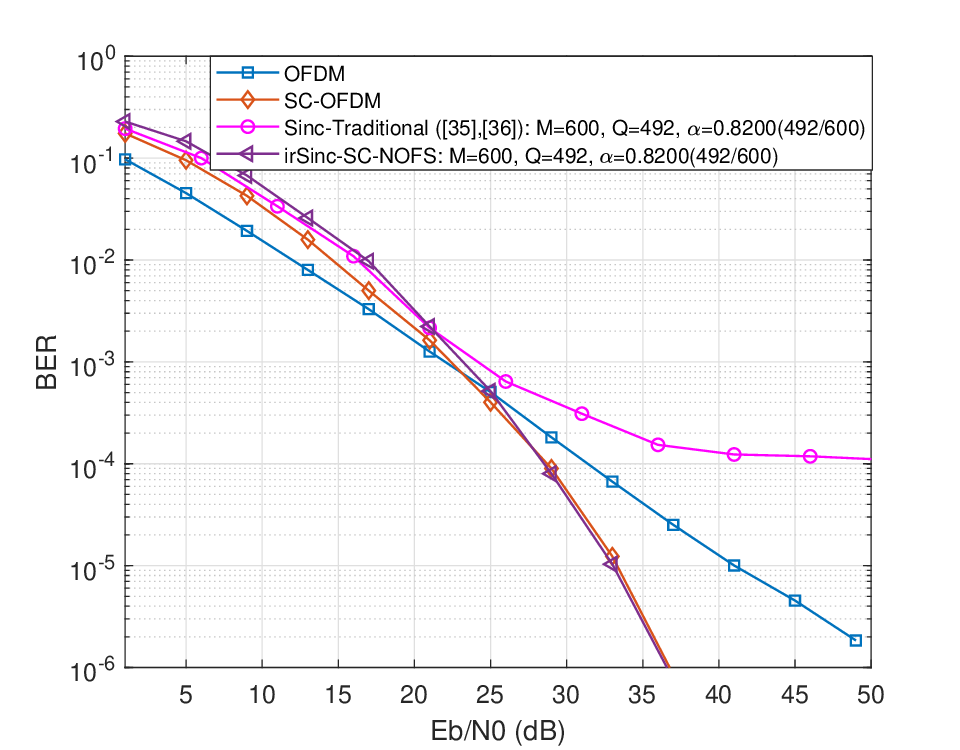}
\end{center}
\caption{{BER performance comparisons in a time-variant frequency selective channel. } }
\label{Fig:BER_channel_SC_NOFS_M600_N1024}
\end{figure}

This work also tests the proposed signals in a time-variant frequency selective channel. The channel's power delay profile is defined by ${h(t)}=0.8765\delta(t)-0.2279\delta(t-T_s)+0.1315\delta(t-4T_s)-0.4032e^{\frac{j\pi}{2}}\delta(t-7T_s)$, with each element following a Rayleigh distribution. The frequency-domain one-tap channel estimation/equalization method is applied with BER demonstrations to show the quality of signals under time-variant multiplath channel effects. It is noted that the channel estimation for the proposed SC-NOFS follows the principle of conventional OFDM and is easily integrated in existing standards. Other non-orthogonal signals need to face complicated and inaccurate channel estimation and equalization challenges while our proposed solution perfectly avoids this. The channel distorted BER results are shown in Fig. \ref{Fig:BER_channel_SC_NOFS_M600_N1024}, in which the Sinc-Traditional(\cite{HC_MCM_2004},\cite{Truncated_OFDM_2016}) signals exhibit the worst performance, particularly at high Eb/N0 ranges due to the signal internal ICI using Sinc shaped sub-carriers. At sufficiently high Eb/N0 levels, the curve tends towards flatness, indicating unrecoverable performance degradation. In the case of traditional OFDM signals, superior performance is observed within low Eb/N0 ranges, showing its resilience when noise power dominates. Both SC-NOFS and SC-OFDM outperform the traditional OFDM due to improved frequency diversity against the multipath propagation impact. It is noted that the SC-NOFS signal exhibits similar performance to SC-OFDM but with 22\% higher spectral efficiency. These results reveal the advantages of SC-NOFS signals over other waveform candidates, even in a time-variant frequency-selective channel.

\subsection{Complexity Comparison} \label{subsec:complexity}

\begin{figure}[t]
\begin{center}
\includegraphics[scale=0.57]{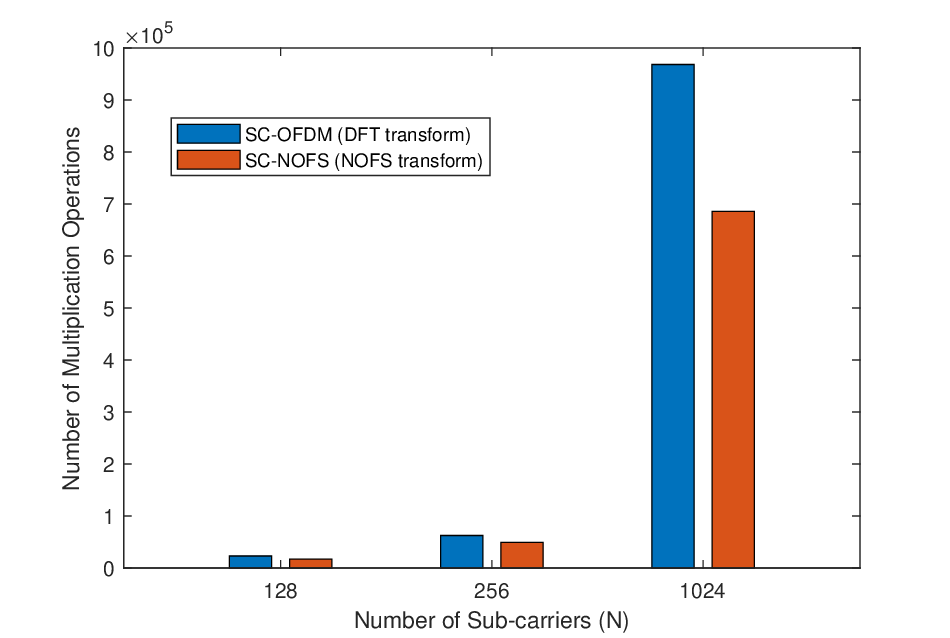}
\end{center}
\caption{Computational complexity in terms of real-value multiplication operations under different numbers of sub-carriers ($N$).  }
\label{Fig:SC_NOFS_mult_complexity}
\end{figure}

In SC-NOFS, a significant innovation is the NOFS transform block and its inverse counterpart, the INOFS transform. Unlike the conventional \ac{DFT} and \ac{IDFT}, illustrated in Fig. \ref{Fig:SC_NOFS_system_block_diagram}(a), where the input and output have the same dimensions, the NOFS transform yields a reduced output dimension, while the INOFS transform features a reduced input dimension. Consequently, the computational complexity in SC-NOFS is lower compared to that of SC-OFDM. Moreover, to fine tune the neural network matrix $\mathbf{W_\gamma}$, unimportant neural connections are pruned, further decreasing the computational load. In contrast, in SC-OFDM, its first stage signal processing relies on the \ac{DFT} where the traditional \ac{FFT} is impossible, leading to the use of direct matrix multiplication, resulting in increased computational complexity. Based on this analysis, our proposed SC-NOFS offers enhanced spectral efficiency while simultaneously reducing computational complexity when compared to SC-OFDM.

In Fig. \ref{Fig:SC_NOFS_mult_complexity}, we test three scenarios with varying numbers of sub-carriers (N=128, 256, 1024). We evaluate the number of real-value multiplications as the metric for computational complexity comparison. It is evident that as the number of sub-carriers increases, so does the demand for multiplications. However, the proposed SC-NOFS demands fewer operations due to the use of NOFS and INOFS transforms.

\section{Hardware Prototyping}\label{sec:hardware_prototyping}

Instead of using high-value equipment \cite{Tongyang_MU_MIMO_NB_IoT_2018, Tongyang_hybrid_precoding_2020, Tongyang_JIOT_2018_double_device, Tongyang_ISAC_OJCS_2022} to validate our proposed communication systems, this work employs the Analog Devices \ac{SDR} PLUTO \cite{PlutoSDR} to test the functionality of our proposed signals. Our experiment is designed to have the following advantages:

\begin{itemize}
\item{{Operation on Low-Cost Hardware: The experiment is conducted using the low-cost Analog Devices SDR PLUTO, enabling cost-efficient \ac{IoT} testing and validation of our proposed signals in IoT preferred devices.}}

\item{{Integration with Standard OFDM: The designed signal is seamlessly integrated into OFDM signals, ensuring compatibility with existing OFDM-supported standards.}}

\item{{Avoidance of AI Model Retraining: Even with changes in channel conditions, there is no need to retrain the AI model, simplifying the experimentation process and demonstrating the feasibility of AI in communication systems with variable channels.}}

\end{itemize}

\begin{figure}[t]
\begin{center}
\includegraphics[scale=0.55]{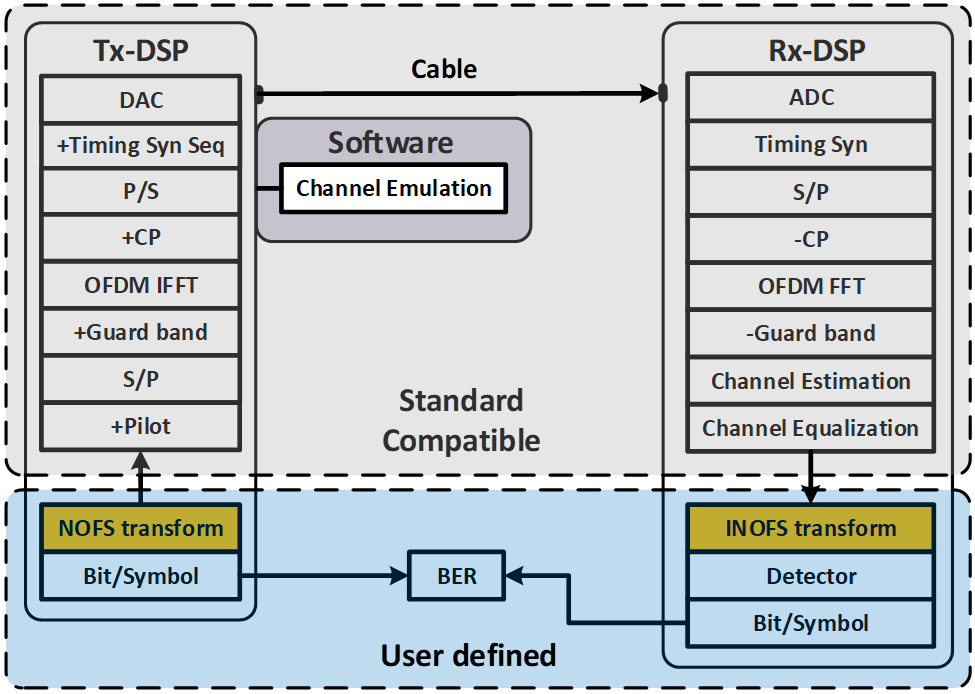}
\end{center}
\caption{Standard compatible SC-NOFS signal experiment block diagram illustration. }
\label{Fig:experiment_SC_NOFS_block_diagram}
\end{figure}

The experiment block diagram is illustrated in Fig. \ref{Fig:experiment_SC_NOFS_block_diagram}. The signal processing blocks within the upper dashed box are standard compatible because the proposed signal framework applies the same signal modulation, demodulation, channel estimation, and channel equalization methods with OFDM standards. The aim is to showcase that our signal is able to be easily integrated into existing standards where simple OFDM signal processing algorithms can be reused. 

In Fig. \ref{Fig:experiment_SC_NOFS_block_diagram}, at the transmitter side, symbols such as \ac{QAM}, are firstly modulated by the NOFS transform. Pilot symbols are inserted for the channel estimation purpose. Guard band is added to reserve a protection gap in frequency-domain. \ac{IFFT} is applied for OFDM signal generation. \ac{CP} is added to avoid \ac{ISI} and later is used to assist channel estimation/equalization. Timing synchronization sequence is added at the beginning of the data to locate the beginning of the data stream at the receiver. Analog signals are obtained after the \ac{DAC}. It is noted that software-based channel effect emulation is included at the Tx-DSP side. At the receiver, after the software emulated channel effects and the wired cable transmission, the channel distorted signal is received and converted to digital signals via the \ac{ADC}. After timing synchronization, OFDM demodulation, channel estimation, and channel equalization, the signals are fed to the INOFS transform block for the second stage symbol recovery. With a signal detector followed, the original symbols are recovered.

\begin{figure}[t]
\begin{center}
\includegraphics[scale=0.45]{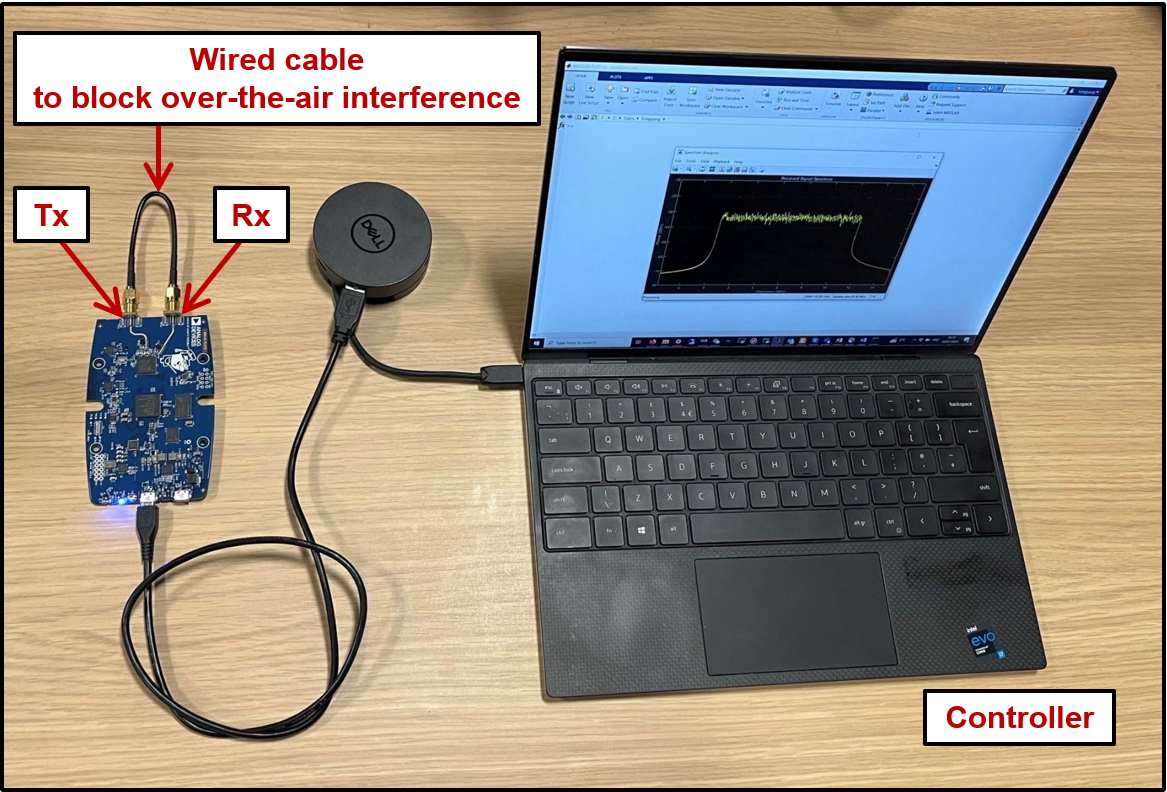}
\end{center}
\caption{Experimental setup at 2.4 GHz carrier frequency using a low-cost SDR device in an indoor lab environment. A wired cable is connected between the Tx port and the Rx port to block over-the-air signal interference.} 
\label{Fig:experiment_SC_NOFS_setup}
\end{figure}

\begin{figure*}[t]
\begin{center}
\includegraphics[scale=0.47]{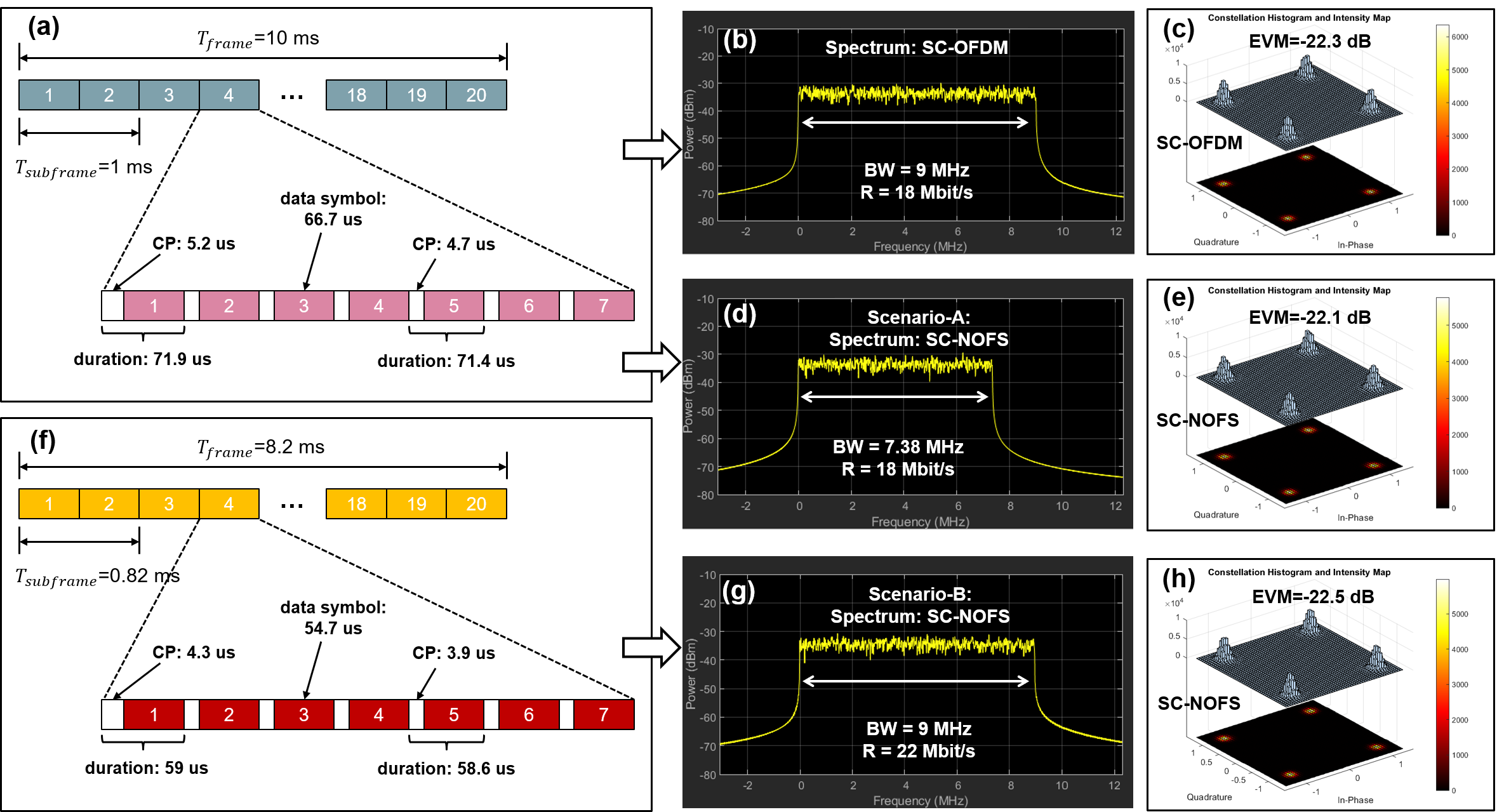}
\end{center}
\caption{The configuration of the PHY-layer time frame and the demonstration of measured spectral and constellation performance. To provide a comprehensive illustration of the entire framework and show standard compatibility, we present a full PHY-layer time frame. In industrial applications, it is worth noting that a shorter packet segment within the time frame could be utilized to align with specific timing requirements. }
\label{Fig:experiment_SC_NOFS_spectrum_constellation}
\end{figure*}

\begin{figure}[t]
\begin{center}
\includegraphics[scale=0.67]{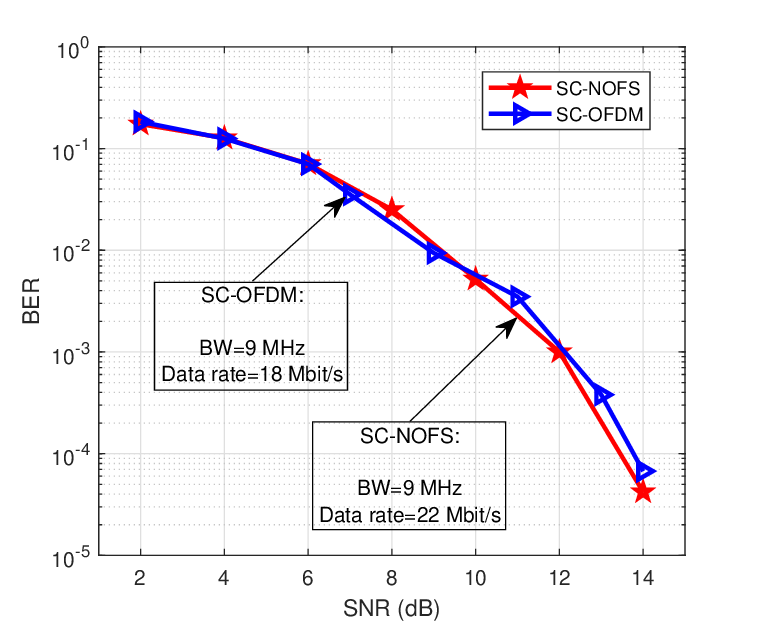}
\end{center}
\caption{Experimental BER performance in the AWGN channel. }
\label{Fig:experiment_BER_SC_NOFS}
\end{figure}

\begin{figure}[t]
\begin{center}
\includegraphics[scale=0.6]{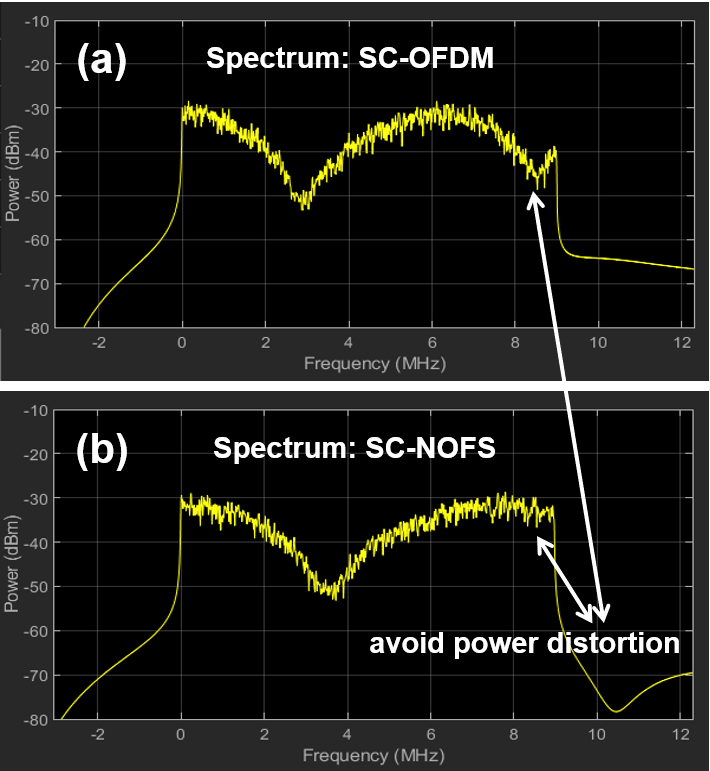}
\end{center}
\caption{Measured frequency selective distortions. }
\label{Fig:experiment_SC_NOFS_spectrum_channel}
\end{figure}

\begin{figure}[t]
\begin{center}
\includegraphics[scale=0.67]{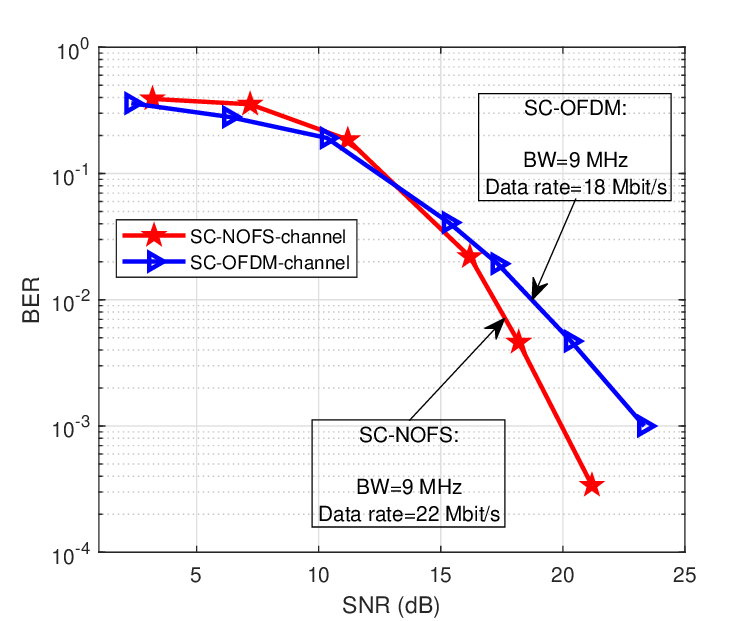}
\end{center}
\caption{Experimental BER performance in a frequency selective channel. }
\label{Fig:experiment_BER_SC_OFDM_NOFS_channel}
\end{figure}

The real hardware setup is shown in Fig. \ref{Fig:experiment_SC_NOFS_setup} where a physical wired cable is connected between the transmitter and the receiver on the \ac{SDR} device. This deliberate choice avoids the influence of over-the-air signal interference, ensuring a controlled environment for experimentation and a convincing and reliable basis for comparisons. To accurately emulate channel effects, the experiment relies on a software emulation methodology, as depicted in Fig. \ref{Fig:experiment_SC_NOFS_block_diagram}. This approach allows for the controlled channel conditions and evaluates their impact on signal transmission, providing valuable insights into system performance under various scenarios. A laptop serves as the central control unit for the entire experimental setup. This setup ensures smooth operation and reliable data collection throughout the experiment. Considering the advantages of SC-NOFS, we design two application scenarios:

\begin{itemize} 
\item{ Scenario-A: Bandwidth saving. As noted in Fig. \ref{Fig:SC_NOFS_system_block_diagram}, the NOFS transform can efficiently prune symbols, indicating that a signal can convey the same information while occupying a reduced spectral bandwidth. This feature is suitable for IoT applications because a large number of IoT devices could be connected in future networks, where potential spectral interference among them is a concern \cite{spectrum_share_2023_spectrum, spectrum_share_2023_PIEEE}. With a narrower data bandwidth and therefore a wider protection guard band, SC-NOFS signals can effectively mitigate such interference. To validate the effectiveness and feasibility of the bandwidth saving scenario, metrics such as spectral utilization, \ac{EVM}, and data rate are measured to provide valuable insights into the performance. } 

\item{ Scenario-B: Faster transmission. By manipulating the signal in Scenario-A via increasing sampling rate, the signal bandwidth expands to that of the conventional SC-OFDM but with a higher data rate. In this case, a system could complete a transmission in a shorter time slot. This leads to a reduction in latency or allowing for the reservation of a wider time protection gap to mitigate severe time jitter effects. To validate the effectiveness and feasibility of the faster transmission scenario, metrics such as spectral utilization, \ac{EVM}, data rate, latency, jitter, and \ac{BER} are measured to quantify the quality and reliability of signal transmission.} 
\end{itemize}

To follow standard specifications such as 5G \cite{Erik_book_5G}, we set the number of data sub-carriers $M$=600, the IFFT size $N$=1024, the sampling rate is 15.36 Mbit/s. The 5G standard defines different subcarrier mapping schemes, and this experiment utilizes a 15 kHz configuration for demonstration purposes. With such configurations, the PHY-layer time frame structure is depicted in Fig. \ref{Fig:experiment_SC_NOFS_spectrum_constellation}(a) where $T_{_{frame}}$=10 ms and each frame contains 10 sub-frames. It is noted that in the experiment we employ a full PHY-layer time frame for illustrations that adhere to standard compatibility. In industrial applications, a shorter packet segment within the time frame could be utilized to meet specific timing requirements. In Fig. \ref{Fig:experiment_SC_NOFS_spectrum_constellation}(b), it is evident that the occupied bandwidth of the SC-OFDM signal is 9 MHz. Utilizing QPSK, the data rate is 18 Mbit/s. With this configuration, the recovered signal constellation is displayed in Fig. \ref{Fig:experiment_SC_NOFS_spectrum_constellation}(c) with \ac{EVM} at -22.3 dB.

In the Scenario-A for SC-NOFS, our proposed signal allows for symbol pruning from 600 to 492, achieving a compression ratio of $\alpha$=0.82. As illustrated in Fig. \ref{Fig:experiment_SC_NOFS_spectrum_constellation}(d), the signal bandwidth is compressed by this ratio to a narrower bandwidth of 7.38 MHz. Despite this reduction in bandwidth due to symbol pruning in our SC-NOFS signal, the receiver can recover the complete set of symbols using specifically designed algorithms. Consequently, the data rate with SC-NOFS remains at 18 Mbit/s. The performance of SC-NOFS is displayed in Fig. \ref{Fig:experiment_SC_NOFS_spectrum_constellation}(e), where four distinct constellation points are clearly observed, with an \ac{EVM} of -22.1 dB, comparable to that of SC-OFDM.

For the Scenario-B using SC-NOFS, a higher sampling rate of 18.73 Mbit/s could be employed. The higher data rate indicates that the system can transmit more data in a given time slot. On the other hand, the system can transmit the same amount of data faster indicating time saving. As illustrated in Fig. \ref{Fig:experiment_SC_NOFS_spectrum_constellation}(f), the duration of one frame is reduced to $T_{_{frame}}$=8.2 ms, with a 1.8 ms protection gap introduced to mitigate the time jitter effect. The spectral bandwidth under the higher sampling rate expands to 9 MHz, similar to that of SC-OFDM, but with a higher data rate at 22 Mbit/s, as shown in Fig. \ref{Fig:experiment_SC_NOFS_spectrum_constellation}(g). Its performance is presented in Fig. \ref{Fig:experiment_SC_NOFS_spectrum_constellation}(h) where four constellation points are clearly observable with an \ac{EVM} of -22.5 dB.

For a more comprehensive analysis of SC-NOFS, it is essential to investigate the \ac{BER} in relation to \ac{SNR}. We introduce noise to the signal in the digital domain to simulate a wide range of SNR conditions. The resulting BER performance is illustrated in Fig. \ref{Fig:experiment_BER_SC_NOFS}. It is evident that both signals exhibit similar performance across different SNR values. One explanation for this similarity lies in the use of OFDM as the air interface for both SC-OFDM and SC-NOFS. As a result, they demonstrate comparable channel estimation and equalization performance. Furthermore, the NOFS transform, having been verified to perform equivalently to \ac{DFT}, contributes to the comparable BER performance observed in both signals.

The use of a wired cable connection eliminates over-the-air signal interference but comes at the cost of losing multipath effects, resulting in flat spectra, as displayed in Fig. \ref{Fig:experiment_SC_NOFS_spectrum_constellation}. To have a robust performance evaluation under multipath effects and show the advantage of SC-NOFS, the experiment specifically designed a frequency-selective channel model in a software environment defined by ${h_B(t)}=0.6965\delta(t)-0.3279\delta(t-T_s)+0.1765\delta(t-3T_s)+0.5991\delta(t-5T_s)+0.1315\delta(t-7T_s)$. By applying the same channel model to SC-OFDM and SC-NOFS, the signals with frequency selective distortions are displayed in Fig. \ref{Fig:experiment_SC_NOFS_spectrum_channel}. It is noted that the faster transmission of SC-NOFS signals helps avoid the power distortion segment of the spectrum in SC-OFDM. This avoidance contributes to the superior BER performance of SC-NOFS compared to SC-OFDM, even with a 22\% higher data transmission rate, as illustrated in Fig. \ref{Fig:experiment_BER_SC_OFDM_NOFS_channel}.

\begin{figure}[t]
\begin{center}
\includegraphics[scale=0.65]{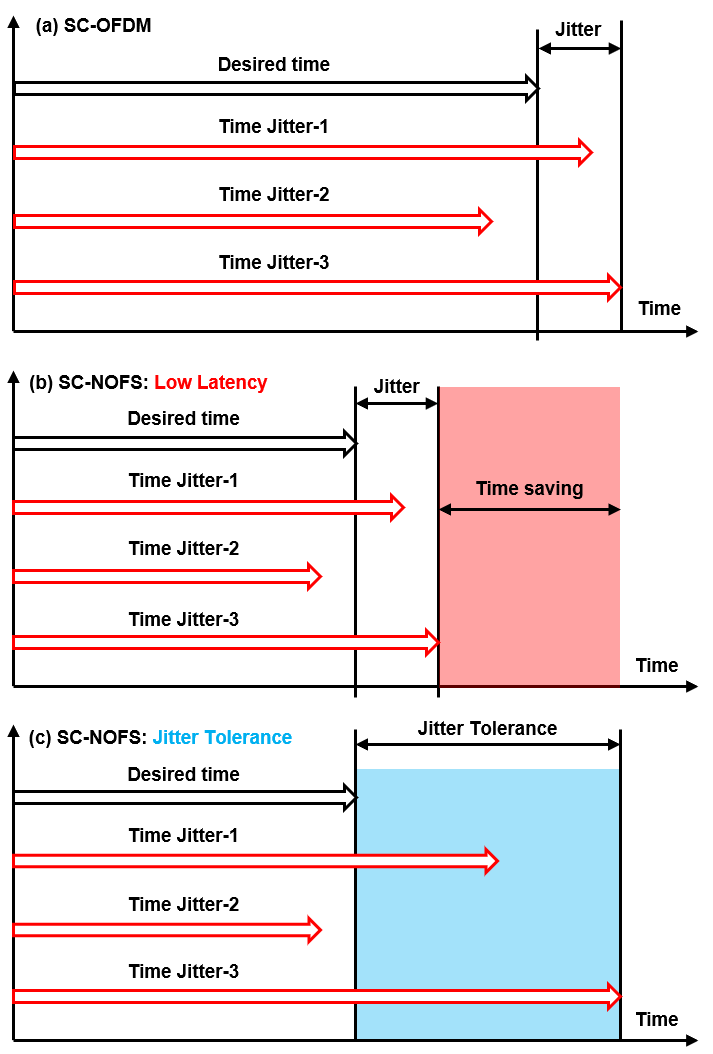}
\end{center}
\caption{Examples showing how SC-NOFS signals reduce transmission latency and enhance time jitter tolerance. }
\label{Fig:SC_NOFS_latency_jitter_block_diagram}
\end{figure}

Considering the timing frame and signal features in Fig. \ref{Fig:experiment_SC_NOFS_spectrum_constellation}(f)(g), SC-NOFS signals offer significant advantages in reducing time latency and addressing challenges associated with time jitter. This makes SC-NOFS a promising candidate for applications where precise timing and minimized variations in signal arrival times are crucial. The advantages of SC-NOFS are displayed in Fig. \ref{Fig:SC_NOFS_latency_jitter_block_diagram} where various time jitter conditions are considered. In Fig. \ref{Fig:SC_NOFS_latency_jitter_block_diagram}(b), the 1.8 ms time saving results in a shorter transmission time, creating a low-latency scenario. Furthermore, the saved time resource can also be utilized to improve time jitter robustness by reserving a wider time gap in Fig. \ref{Fig:SC_NOFS_latency_jitter_block_diagram}(c), thereby enhancing jitter tolerance.

\section{Conclusions}\label{sec:conclusion}

This work proposed a spectrally efficient SC-NOFS waveform enabled by irSinc to enhance performance in industrial IoT applications. Specifically, we propose a two-stage signal waveform framework that aligns with existing 5G standards. The framework leverages the dimension-reduced NOFS transform, which has higher spectral efficiency and lower computational complexity compared to DFT. Furthermore, the framework employs machine learning to determine the optimal irSinc sub-carriers rather than the signal itself. As a result, the AI-based communication framework is robust, eliminating the need for model re-training even with changing channels. The signal achieves a spectral compression of $\alpha$=0.82, indicating a 22\%=(1/$\alpha$)$\times$100\% improvement in spectral efficiency. Simulation results demonstrate that the SC-NOFS signal achieves a better PAPR and competitive BER in both AWGN and frequency-selective channels. Experimental validation on a low-cost SDR device with practical signal transmissions confirms two scenarios in a 5G setup: 1) Bandwidth saving. Conventional SC-OFDM requires a 9 MHz bandwidth for an 18 Mbit/s data rate, while the proposed SC-NOFS only requires 7.38 MHz for the same data rate. 2) Faster transmission. Utilizing the same spectral bandwidth of 9 MHz, SC-OFDM provides an 18 Mbit/s data rate, while SC-NOFS achieves 22 Mbit/s. In a 5G setup, the faster transmission saves 1.8 ms of timing resources, leading to either reduced latency or enhanced time jitter tolerance.

\bibliographystyle{IEEEtran}
\bibliography{Ref_SC_NOFS}

\end{document}